\documentclass[10pt]{article}
\usepackage[french,english]{babel}
\usepackage{amsmath}
\usepackage{amsfonts}
\usepackage{amssymb}
\usepackage{graphicx}                              
\usepackage{times}
\usepackage{version}     
\usepackage[makeroom]{cancel} 
\begin{document}

%
December 1st 2017 \hfill 
\vskip 5cm
{\baselineskip 18pt
\begin{center}
{\bf 1-LOOP MASS GENERATION BY A CONSTANT EXTERNAL MAGNETIC FIELD
FOR AN ELECTRON PROPAGATING IN A THIN MEDIUM
}
\end{center}
}
\baselineskip 16pt
\arraycolsep 3pt  
%
\vskip .2cm
\centerline{
B.~Machet
     \footnote{Sorbonne Universit\'es, UPMC Univ Paris 06, UMR 7589,
LPTHE, F-75005, Paris, France}
     \footnote{CNRS, UMR 7589, LPTHE, F-75005, Paris, France.}
     \footnote{Postal address:
LPTHE tour 13-14, 4\raise 3pt \hbox{\tiny \`eme} \'etage,
          UPMC Univ Paris 06, BP 126, 4 place Jussieu,
          F-75252 Paris Cedex 05 (France)}
    \footnote{machet@lpthe.jussieu.fr}
     }
\vskip 1cm

{\bf Abstract:} The 1-loop self-energy of a Dirac electron of mass $m$
 propagating in a thin medium simulating graphene in an external magnetic field $B$
 is investigated in Quantum Field Theory.
Equivalence is shown with the so-called reduced QED$_{3+1}$ on a 2-brane.
Schwinger-like methods are used to calculate the self-mass $\delta m_{LLL}$ of the electron
 when it lies in the lowest Landau level.
Unlike in standard QED$_{3+1}$, it does not vanish at the  limit $m\to 0$ :
$\delta m_{LLL} \stackrel{m\to 0}{\to} \displaystyle\frac{\alpha}{2}\;
\sqrt{\displaystyle\frac{\pi}{2}}\,\sqrt{\displaystyle\frac{\hbar\,|e|B}{c^2}}$
(with $\alpha=\displaystyle\frac{e^2}{4\pi\hbar c}$);
all Landau levels of the  virtual electron are taken
into account and on mass-shell renormalization conditions are implemented.
Restricting to the sole  lowest Landau level of the
virtual electron is explicitly shown to be inadequate.
Resummations at higher orders lie beyond the scope of this work.

\bigskip


\newpage
%
%
\section{Introduction}

Graphene is known as a quasi 2+1 dimensional medium with Dirac-like
massless electrons (a gapless medium) -- see for example
\cite{CastroNeto2009}.
 Whether or not and in which
circumstances a gap can  open has important consequences, for example
on electrical and optical properties \cite{Goerbig2011}.
 In addition to purely theoretical
motivations, this is one of the reasons why we study in this work the
spontaneous emergence of a gap for a model which can naively mimic
graphene in the presence of a constant external
magnetic field $B$.

While it is generally admitted that the presence of  $B$ is likely to
trigger chiral symmetry breaking (see for example
\cite{MiranskyShovkovy2015}), the demonstrations usually rely on
various approximations. In standard QED$_{3+1}$, they are often
based on the dimensional reduction that operates in the presence of a
strong $B$ \cite{LS1979} \cite{GusyninMiranskyShovkovy1996}
 and on resummations of a certain class of diagrams
\cite{LS1981} \cite{KMO-MPLA-2002} (which become
suspicious after realizing that only double logs have been taken into
account, leaving aside large single logs \cite{Machet1}).
Also, various approximations to coupled Dyson-Schwinger equations
are invoked, associated to the use of very special gauges to simplify the
vertex (see \cite{GusyninMiranskyShovkovy1999}); this makes the
demonstrations
tedious, not very transparent and possibly controversial.
In reduced QED$_{3+1}$ on a 2-brane, which is often considered to provide a
fair
description of graphene, other approximations are invoked, like the
dominance of
the lowest Landau level
\cite{GusyninMiranskyShovkovyNPB1999} while it was shown, for example in
\cite{KMO-MPLA-2002}, that higher levels are important and trigger charge
renormalization;
moreover the language that is used is often confusing for people working in
Quantum Field Theory.

The calculation of the 1-loop self-energy of an electron propagating in an
external $B$ that I present here uses
the sole techniques of Quantum Field Theory.
The external electron is chosen, for the sake of simplicity,
 to lie  in the lowest Landau level (LLL), and, in
this case, analytical (quasi-)exact formul{\ae} can be obtained
by using the formalism of Schwinger \cite{Schwinger} as it is carefully
explained in \cite{DittrichReuter}.

I previously tackled the case of standard QED$_{3+1}$ in \cite{Machet1} by calculating the
integral of Demeur \cite{Demeur} and Jancovici \cite{Jancovici}
 beyond the leading $\big(\ln \frac{|e|B}{m^2}\big)^2$ approximation.
I demonstrated that large logarithms had
been overlooked and, then, neglected; they are tightly connected with the
counterterms needed to implement suitable renormalization conditions.
In this case, $\delta m \to 0$ when $m \to 0$.

These calculations are adapted here  to a thin graphene-like medium.
They are explained step by step such that they should appear fairly easy to
reproduce, with no obscure gap to fill. They mostly go along the lines of
\cite{DittrichReuter}, and differences are outlined.
A massive Dirac
electron is considered to  propagate inside a thin film of thickness $2a$, 
the Hamiltonian of which being deprived of its ``$p_3 \gamma_3$'' term
(see for example \cite{Goerbig2011}).  $B$, supposed to be
static and uniform is  considered to be directed along the $z$
axis orthogonal to the medium strip. To make the calculation simpler
and more transparent, no Fermi velocity different from the
speed of light is introduced, such that I will be dealing with a special
avatar of Quantum Electrodynamics, and extra degeneracies present in graphene
\cite{Goerbig2011} are eluded.
The topic of symmetries  will not be dwelt on either
(see the review \cite{MiranskyShovkovy2015} on this subject).

As I will demonstrate by working in position space, 
this model yields for the electron self-energy the same
expression as reduced QED$_{3+1}$ on a 2-brane \cite{Gorbar2001}
\cite{Pevzner2009}: the effective photon propagator
turns out, indeed, to be the one of standard QED$_{3+1}$ integrated over
its $k_3$ momentum.
For the internal electron propagator in presence of an
external $B$ I  use Schwinger's \cite{Schwinger1949} and Tsai's
\cite{Tsai1974} expression, which accounts for all Landau levels,
 adapted to the particular situation and Hamiltonian under consideration.
The calculations are (and should)
 be performed with a non-vanishing electron mass $m$ before the limit
$m\to 0$ is taken.
In the last part I only take into account the LLL of the internal
electron, and show that neglecting higher levels is a bad approximation.

To avoid confusion, let me stress  that all spinors and $\gamma$
matrices that are considered in this work  are 4-dimensional.
Any eventual connection with QED$_{2+1}$,
if any, can accordingly only be quite remote, and we shall not dwell on this any
more.


So, though  the result that I exhibit will certainly not be a surprise
for many, I hope that the rigorous demonstration of a simple and  exact
formula that anyone can check with standard techniques will bring 
$B$-triggered mass generation  from radiative corrections on a more solid ground.
Like for QED$_{3+1}$, renormalization conditions and  the counterterms that
must be introduced to fulfill them play important roles
\footnote{In the work \cite{Machet2} I emphasize their role in the
calculation of the photon vacuum polarization for the same graphene-like
medium as the one considered here.}.

A major challenge is also, there, to deal with a strongly coupled
theory since a 1-loop result is certainly meaningless when the coupling
constant gets of order $1$. The necessary resummations
look  highly non-trivial since they  do not only concern
double and / or simple logs, but more complicated functions, and
they have furthermore, of course, to be performed while satisfying at each
order appropriate renormalization conditions. To my knowledge this last
requirement has never been satisfied  and
tackling such formidable tasks lies largely beyond the scope of this work.

\section{Propagation inside a thin, graphene-like medium; equivalence with
reduced QED$\boldsymbol{_{3+1}}$ on a 2-brane}
\label{section:propa}

A general argumentation concerning reduced QED can be found, for example,
 in \cite{Gorbar2001}.
A more down-to-earth determination of the effective photon
propagator is nevertheless instructive because it provides
a simpler understanding of the mechanisms at work, and also because 
this approach can be applied  to vacuum polarization \cite{Machet2},
yielding less-trivial results.

Let us calculate in  position space
the electron propagator $G(y,x)$ at 1-loop depicted in Fig.~1 (including
external legs). We
call $G_0$ the tree-level electron propagator in the presence of $B$
\footnote{The results of this paragraph do not depend whether the external
$B$ is present or not.}
(described by the double lines in Fig.~1) and
$\Delta_{\mu\nu}$ the bare photon propagator. 

\begin{center}
\includegraphics[width=7 cm, height=3 cm]{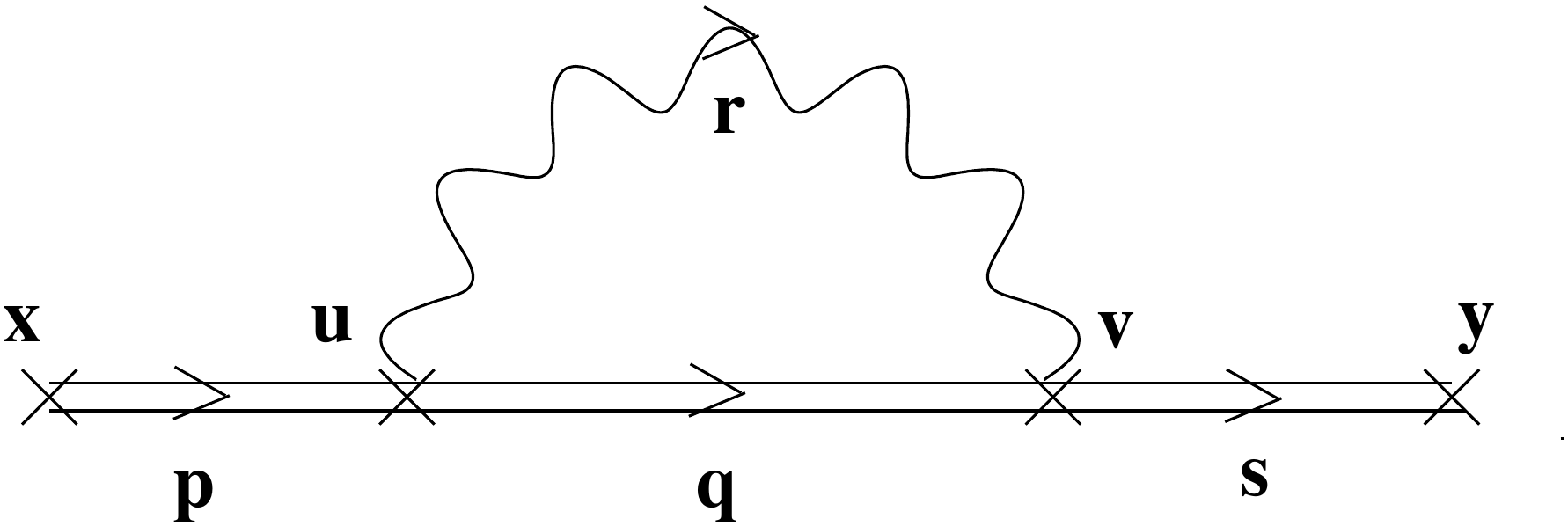}

\smallskip
{\em Fig.~1: the 1-loop electron propagator in external $B$}
\end{center}
 
One has
\begin{equation}
G(x',x'') \equiv i<0\;|\;T\;\psi(x')\bar\psi(x'')\;|\;0> =\Phi(x',x'')
\int\frac{d^4p}{(2\pi)^4}\;e^{ip(x''-x')}G(p)
\label{eq:genprop0}
\end{equation}
in which the phase \cite{Tsai1974}
\begin{equation}
\Phi(x',x'')=e^{i|e|q\int_{x''}^{x'} dx_\mu A^\mu(x)}
\label{eq:phase}
\end{equation}
ensures the gauge invariance of the Green function
($A^\mu$ is the vector potential).

To avoid confusion, the unit of electric charge we note $|e|$ such
that the electron charge is $-|e|$. 
In \cite{Schwinger} and in \cite{Tsai1974}, this unit of electric charge is
instead noted $e$. In \cite{DittrichReuter}, $e$ denotes instead  the (negative)
electron charge. We shall see that these precisions are important, in
particular to get the appropriate propagator for the LLL of an electron.

Like in \cite{Schwinger} and \cite{Tsai1974} we  introduce $q$ such
that $q|e|$ is the electron charge. Therefore $q=-1$
\footnote{This $q$ should not be confused with the 4-momentum that appears in Fig.~1. The reader will easily make the difference.}. This makes the covariant derivative  $D_\mu =\partial_\mu -i|e|q A_\mu$ such that
$\pi_\mu =\displaystyle\frac{1}{i}\partial_\mu = p_\mu +|e|A_\mu$.

For any 4-vector $v=(v_0,v_1,v_2,v_3)$, it is useful to introduce the
notations $\hat v=(v_0,v_1,v_2,0)$, $v_\parallel=(v_0,0,0,v_3)$ and
$v_\perp=(0,v_1,v_2,0)$.

The 1-loop electron propagator depicted in Fig.~1 writes
\begin{equation}
\begin{split}
iG(y,x) 
%
&= -e^2\; e^{i|e|q\int_y^x dt_\mu A^\mu(t)}
\int d^4u\int d^4v\;
\int\frac{d^4 p}{(2\pi)^4}\;e^{ip(u-x)}\;G_0(p)
\int\frac{d^4 q}{(2\pi)^4}\;e^{iq(v-u)}\;G_0(q)\cr
& \hskip 2cm \gamma^\mu
\int\frac{d^4r}{(2\pi)^4}\;e^{ir(v-u)}\;\Delta_{\mu\nu}(r)
\gamma^\nu
\int\frac{d^4 s}{(2\pi)^4}\;e^{is(y-v)}\;G_0(s).
\end{split}
\end{equation}

We now specialize to the medium under concern and
 consider ``graphene-like'' electrons propagating inside a thin film of
thickness $2a$.
This situation has two consequences:\newline
*\ $G_0(q)=G_0(\hat q),\quad G_0(p)=G_0(\hat p),\quad G_0(s)=G_0(\hat s)$ get
deprived of their $\gamma_3$ components;\newline
*\ the vertices at which the electron and photon interact
  being located inside the strip, the integrals on
their positions $u_3$ and $v_3$ along the $z$ axis should be truncated to
$\int_{-a}^{+a}du_3\int_{-a}^{+a}dv_3$. This gives
\begin{equation}
\begin{split}
iG(y,x)
&=-e^2\; e^{i|e|q\int_y^x dt_\mu A^\mu(t)}
\int\frac{dq_3}{2\pi} \int\frac{dp_3}{2\pi} \int\frac{dr_3}{2\pi}
\int\frac{ds_3}{2\pi}
\int_{-a}^{+a} du_3\;e^{iu_3(p_3-q_3-r_3)}\int_{-a}^{+a}
dv_3\;e^{iv_3(q_3+r_3-s_3)}\cr
&  \hskip 1cm\int\frac{d^3\hat p}{(2\pi)^3}\;e^{i\hat p(\hat y -\hat x)-ip_3x_3+is_3y_3}
G_0(\hat p)
\int\frac{d^3\hat q}{(2\pi)^3}\;G_0(\hat q)\gamma^\mu \Delta_{\mu\nu}(\hat
p-\hat q,r_3)\gamma^\nu\;G_0(\hat p).
\end{split}
\end{equation}
The two integrations $\int du_3$ and $\int dv_3$ can be performed since
\begin{equation}
\int_{-a}^{+a}dx\;e^{itx}= \frac{2\sin at}{t},
\end{equation}
which leads to
\begin{equation}
\begin{split}
iG(y,x) &=-4 e^2\; e^{i|e|q\int_y^x dt_\mu A^\mu(t)}
\int\frac{dp_3}{2\pi} \int\frac{dr_3}{2\pi}
\int\frac{ds_3}{2\pi}
\Big[\int\frac{dq_3}{2\pi}\;
\frac{\sin a(q_3+r_3-s_3)}{q_3+r_3-s_3}\;\frac{\sin
a(p_3-q_3-r_3)}{p_3-q_3-r_3}\Big]\cr
&\hskip 1cm \int\frac{d^3\hat p}{(2\pi)^3}\;e^{i\hat p(\hat y -\hat
x)-ip_3x_3+is_3y_3} G_0(\hat p)
\int\frac{d^3\hat q}{(2\pi)^3}\;G_0(\hat q)\gamma^\mu \Delta_{\mu\nu}(\hat
p-\hat q,r_3)\gamma^\nu\;G_0(\hat p).
\end{split}
\end{equation}
In there the integration $\int dq_3$ can also be done explicitly since
\begin{equation}
\int dq_3\;\frac{\sin a(q_3-\sigma)}{q_3-\sigma}\frac{\sin
a(q_3-\tau)}{q_3-\tau}=\pi\;\frac{\sin a(\sigma-\tau)}{\sigma-\tau},
\end{equation}
with $\sigma=s_3-r_3, \tau=p_3-r_3$, which has the property to be independent of
$r_3$. We get now
\begin{equation}
\begin{split}
iG(y,x) &= -2e^2\; e^{i|e|q\int_y^x dt_\mu A^\mu(t)}
\int\frac{dp_3}{2\pi}\int\frac{ds_3}{2\pi}
\frac{\sin a(s_3-p_3)}{s_3-p_3}\int\frac{d^3\hat p}{(2\pi)^3}\;e^{i\hat
p(\hat y -\hat x)-ip_3x_3+is_3y_3}
G_0(\hat p)\cr
& \hskip 3cm \int\frac{d^3\hat q}{(2\pi)^3}\int\frac{dr_3}{2\pi}
\;G_0(\hat q)\gamma^\mu \Delta_{\mu\nu}(\hat p-\hat
q,r_3)\gamma^\nu\;G_0(\hat p).
\end{split}
\end{equation}
Going to the new variables $h_3=s_3+p_3,l_3=s_3-p_3 \Rightarrow dp_3\, ds_3
= \frac12 dh_3\, dl_3$ yields
\begin{equation}
\begin{split}
iG(y,x) &= -e^2\; e^{i|e|q\int_y^x dt_\mu A^\mu(t)}
\int\frac{dl_3}{2\pi}\;\frac{\sin al_3}{l_3} \;e^{il_3\frac{x_3+y_3}{2}}
\int \frac{dh_3}{2\pi} e^{ih_3\frac{y_3-x_3}{2}}
\int\frac{d^3\hat p}{(2\pi)^3}\;e^{i\hat p(\hat y -\hat x)}G_0(\hat p)\cr
& \hskip 2cm \int\frac{d^3\hat q}{(2\pi)^3}\int\frac{dr_3}{2\pi}
\;G_0(\hat q)\gamma^\mu \Delta_{\mu\nu}(\hat p-\hat
q,r_3)\gamma^\nu\;G_0(\hat p).
\end{split}
\end{equation}
The condition $x_3 + y_3 \leq 2a$ is verified because the electrons are
constrained to propagate inside the strip, such that
\begin{equation}
\int\frac{dl_3}{2\pi}\;\frac{\sin al_3}{l_3} \;e^{il_3\frac{x_3+y_3}{2}}
=\frac12.
\end{equation}
This yields
\begin{equation}
iG(y,x)=-e^2
\int\frac{d^4 p}{(2\pi)^4}\;
e^{i p(y - x)}G_0(\hat p)
\; e^{i|e|q\int_y^x dt_\mu A^\mu(t)}
\int\frac{d^3\hat q}{(2\pi)^3}\int\frac{dr_3}{2\pi}
\;G_0(\hat q)\gamma^\mu \Delta_{\mu\nu}(\hat p-\hat q,r_3)\gamma^\nu
G_0(\hat p).
\end{equation}
The self-energy $\Sigma$ is obtained from the 1-loop propagator
 above by chopping
off the two external fermion $iG_0$ propagators, which leads to
\begin{equation}
\Sigma(x,y)=\Phi(x,y)\int\frac{d^4p}{(2\pi)^4}\;
e^{ip(y-x)}\;\Sigma(\hat p),
\label{eq:self0}
\end{equation}
with the phase $\Phi$ given in (\ref{eq:phase}) and to
\begin{equation}
i\Sigma(\hat p) = e^2\int\frac{d^3\hat
k}{(2\pi)^3}\int\frac{dr_3}{2\pi}
\;G_0(\hat p-\hat k)\gamma^\mu \Delta_{\mu\nu}(\hat k,r_3)\gamma^\nu,
\end{equation}
in which, to avoid conflicts between notations, we have made the change of variables $p-q 
\to k$ in the momenta, which amounts to labeling them like in \cite{DittrichReuter}.

This shows the equivalence with reduced QED$_{3+1}$ on a 2-brane, in which
the ``effective'' internal photon propagator is (see \cite{Gorbar2001})
\begin{equation}
\tilde\Delta_{\mu\nu}(\hat k)= \int \frac{dr_3}{2\pi}\;
\Delta_{\mu\nu}(\hat k, r_3).
\label{eq:photoneff}
\end{equation}
In the Feynman gauge
\footnote{The choice of a special gauge is of course not optimal but is
justified by the property that the formalism of Schwinger is gauge
invariant \cite{Schwinger1949}.}
 one gets
$\tilde\Delta_{\mu\nu}(\hat k)
=\displaystyle\int\displaystyle\frac{dr_3}{2\pi}\;\frac{g_{\mu\nu}}{\hat k^2 +r_3^2}
=\displaystyle\frac{1}{2}\;\displaystyle\frac{g_{\mu\nu}}{\sqrt{\hat k^2}}$ such that
\begin{equation}
i\Sigma(\hat p)=-\frac{e^2}{2} \int\frac{d^3\hat k}{(2\pi)^3}\;\gamma^\mu\;
G_0(\hat
p-\hat k)\; \frac{g_{\mu\nu}}{\sqrt{\hat k^2}}\; \gamma^\nu.
\label{eq:brane}
\end{equation}
which should be compared with eq.~(3.9) of \cite{DittrichReuter}.

No dependence on the thickness $a$ of the
medium occurs anymore (unlike for the vacuum polarization \cite{Machet2}).
This is easily understood since we constrained the fermion to propagate inside the
medium (while, for the vacuum polarization, the photon is allowed to also propagate
in the ``bulk'').

\section{The self-energy and self-mass of the electron}

In the whole paper,
we use the metric $(-\;+\;+\;+)$ like in \cite{Schwinger}, \cite{Tsai1974}
and \cite{DittrichReuter}.

The conventions for $\gamma$ matrices and Pauli $\vec\sigma$ matrices are
the same as in \cite{Tsai1974},
\cite{DittrichReuter} and \cite{Schwinger}. In particular,
$\{\gamma^\mu,\gamma^\nu\}=-2g^{\mu\nu}$. We shall denote
(abusively) $\sigma^3 \equiv\sigma^{12}=\frac{1}{2}[\gamma^1,\gamma^2] =
diag(1,-1,1,-1)$; it should not be mistaken for the corresponding $2\times
2$ Pauli matrix.

With these conventions, for an external magnetic field $B$ along the $z$ axis,
 the wave function of the lowest Landau level $|LLL>$ is proportional to
$\left(\begin{array}{c} 0\cr 1\cr 0\cr 0\end{array}\right)$
(\cite{Luttinger} \cite{KuznetsovMikheev}) such that
$\sigma^3 |LLL> =(-1)|LLL>$ and $(1-i\gamma^1 \gamma^2)|LLL> \equiv
(1-\sigma^3)|LLL> = 2|LLL>$.

\subsection{The self-energy in momentum space}

We now proceed to calculating  the self-energy expressed in (\ref{eq:brane}),
following the procedure given in \cite{DittrichReuter}. To that purpose, we
introduce 2 Schwinger parameters: $s_2$ for the photon and $s_1$ for the
electron.

As far as  the photon is concerned, 
instead of $\frac{1}{k^2-i\epsilon}=i\int_0^\infty
ds_2\;e^{-is_2(k^2-i\epsilon)}$ (eq.~(3.10) of \cite{DittrichReuter}),
that is used to represent the 4-dimensional photon propagator in the Feynman gauge,
 we shall use now, according to (\ref{eq:photoneff}) and (\ref{eq:brane})
\begin{equation}
\frac{1}{\sqrt{\hat k^2 -i\epsilon}}=\sqrt{\frac{i}{\pi}}\int_0^\infty
ds_2\; \frac{e^{-is_2(\hat k^2-i\epsilon)}}{\sqrt{s_2}}.
\label{eq:trick}
\end{equation}
However, it is important (see just above (\ref{eq:sig00})) to
use Tsai's \cite{Tsai1974} formul{\ae} and not the ones used in
\cite{DittrichReuter}.

As for the electron, in general
QED$_{3+1}$, its propagator is given (see eq.~(6) of \cite{Tsai1974})
by
\begin{equation}
G_0(k,B)=i\int_0^\infty ds_1\;e^{-is_1\big(m^2-i\epsilon +k_\parallel^2
+\frac{\tan z}{z} k_\perp^2 \big)}
\; \frac{e^{i qz\sigma^3}}{\cos z}\Big(m-k\!\!\!/_\parallel
-\frac{e^{-iqz\sigma^3}}{\cos z} k\!\!\!/_\perp \Big),
\quad z=|e|Bs_1,
\label{eq:ferprop}
\end{equation}
and, in position space by  equations similar to  (\ref{eq:genprop0}) and (\ref{eq:phase}).
These expressions only need to be trivially adapted to the ``truncated''
momenta $\hat p$ and $\hat k$ (see section \ref{section:propa}).

As shown in appendix \ref{section:LLLprop}, (\ref{eq:ferprop}) leads to the adequate propagator
for the $LLL$ at the limit $B \to \infty$. It is in particular proportional
to the customary projector $1-i\gamma^1\gamma^2$, This is not the case of
eq.~(2.47b) of \cite{DittrichReuter} (in there $e<0$), which involves
$e^{i\sigma^3 z}$ instead of $e^{iq\sigma^3 z}$ and leads to the
wrong projector $1+i\gamma^1\gamma^2$ and, later, to confusions and
problems.

From (\ref{eq:brane}) and using (\ref{eq:trick}) and (\ref{eq:ferprop}),
 one gets
instead of (3.11) of \cite{DittrichReuter} (``c.t.'' means ``counter
terms)
\begin{equation}
\begin{split}
\Sigma(\hat p) &= -\sqrt{\frac{i}{\pi}}\;\frac{e^2}{2}\int_0^\infty ds_1
\int_0^\infty\frac{ds_2}{\sqrt{s_2}}\;
\int\frac{d^3\hat k}{(2\pi)^3}\;e^{-is_2(\hat k^2-i\epsilon)}
 e^{-is_1\big(m^2+(\hat p-\hat k)_\parallel^2 + \frac{\tan z}{z}(\hat p-\hat
k)_\perp^2\big)}\cr
& \gamma^\mu\;\frac{e^{iqz\sigma^3}}{\cos z}
\Big[m-(\hat p\!\!\!/-\hat k\!\!\!/)_\parallel - \frac{e^{-iqz\sigma^3}}{\cos z}
(p\!\!\!/-k\!\!\!/)_\perp\Big]\gamma_\mu + c.t.,\quad with\  z=|e|Bs_1,
\end{split}
\label{eq:sig00}
\end{equation}
Since the Hamiltonian of the Dirac electron is presently considered to
be deprived of its $\gamma_3(p-k)_3$ part,
 $(\hat p-\hat k)_\parallel^2 = -(p_0-k_0)^2,
\quad (\hat p\!\!\!/-\hat k\!\!\!/)_\parallel = -\gamma_0(p_0-k_0)$, while
preserving
$(\hat p-\hat k)_\perp^2=(p_1-k_1)^2 + (p_2-k_2)^2$ and
$(p\!\!\!/-k\!\!\!/)_\perp=\gamma_1(p_1-k_1)+\gamma_2(p_2-k_2)$.

One performs the same  change of variable as (3.12) of \cite{DittrichReuter}
\begin{equation}
s_1=su,\quad s_2 = s(1-u) \Rightarrow ds_1\;\frac{ds_2}{\sqrt{s_2}}=
ds\;\sqrt{s}\;\frac{du}{\sqrt{1-u}},
\label{eq:changevars}
\end{equation}
and one still introduces  $y=|e|Bsu$.

The exponentials are then re-expressed in view of performing the $\int d^3
\hat k$ integration. Following a procedure identical to that in
\cite{DittrichReuter} yields, instead of their (3.17)
\begin{equation}
\begin{split}
\Sigma(\hat p) &= -i\frac{e^2}{2}\sqrt{\frac{i}{\pi}}
\int_0^\infty ds\sqrt{s}\;\int_0^1 \frac{du}{\sqrt{1-u}}\;\frac{1}{\cos y}
\Bigg\{\int\frac{d^3\hat k}{(2\pi)^3}\;e^{-is\chi}\Bigg\}\cr
& \gamma^\mu e^{iqy\sigma^3}\Big[
m-(1-u)p\!\!\!/_\parallel+\frac{e^{-iqy\sigma^3}}{\cos y}
\frac{1-u}{1-u+u\tan y/y}p\!\!\!/_\perp
\Big]\gamma_\mu + c.t.,
\end{split}
\label{eq:sig4}
\end{equation}
in which $\chi$ and $\varphi$ are still given by (3.14), (3.15) of
\cite{DittrichReuter}
\begin{equation}
\begin{split}
\chi &= um^2+\varphi+(k_\parallel-up_\parallel)^2 +\big(1-u+u\frac{\tan y}{y}\big)
\Big[k_\perp-\frac{u\tan y/y}{1-u+u\tan y/y}\;p_\perp\Big]^2,\cr
\varphi &= u(1-u)\,p_\parallel^2+\frac{u}{y} \frac{(1-u)\sin y}{(1-u)\cos
y+u\sin y/y}\;p_\perp^2.
\end{split}
\label{eq:chifi}
\end{equation}
The shifts in the integration variables are naturally $k_\parallel \to
k_\parallel - up_\parallel$ and $k_\perp \to k_\perp - -\frac{u\tan
y/y}{1-u+u\tan y/y}\;p_\perp$.

One has to redo the $k$ integrations (which only concerns the integral
inside curly brackets in (\ref{eq:sig4})) since it is now $\displaystyle\int\frac{d^3\hat
k}{(2\pi)^3}$ instead of $\displaystyle\int\frac{d^4 k}{(2\pi)^4}$ for standard
QED$_{3+1}$.
This is simple with the aid of
the standard integral 
\begin{equation}
\displaystyle\int_{-\infty}^{+\infty} dx\;e^{\pm i A x^2}=e^{\pm
i\pi/4}\left(\displaystyle\frac{\pi}{A}\right)^{1/2},\; A>0,
\label{eq:standint}
\end{equation}
and leads to
\begin{equation}
\begin{split}
\Sigma(\hat p) 
&= -\frac{me^2}{16\pi^2}\int_0^\infty
\frac{ds}{s}\int_0^1\frac{du}{\sqrt{1-u}}\;
\frac{e^{-is(um^2+\varphi)}}{(1-u)\cos y+u\sin y/y}\cr
&\hskip 1.5cm \gamma^\mu e^{iqy\sigma^3}\Big[
1-(1-u)\frac{p\!\!\!/_\parallel}{m}+\frac{e^{-iqy\sigma^3}}{\cos y}
\frac{1-u}{1-u+u\tan y/y}\frac{p\!\!\!/_\perp}{m}
\Big]\gamma_\mu + c.t.
\end{split}
\end{equation}
It is then simple matter to perform the Dirac algebra,
which leads, instead of  eq.~(3.27) of \cite{DittrichReuter}, to
\begin{equation}
\begin{split}
\Sigma(\hat p) &= \frac{\alpha m}{4\pi}\int_0^\infty
\frac{ds}{s}\int_0^1\frac{du}{\sqrt{1-u}}\;
\frac{e^{-is(um^2+\varphi)}}{(1-u)\cos y+u\sin y/y}\;e^{iqy\sigma^3}\cr
& \hskip 2cm \Big[1+ e^{-2iqy\sigma^3}+(1-u)e^{-2iqy\sigma^3}
\frac{p\!\!\!/_\parallel}{m} +(1-u)\frac{e^{-iqy\sigma^3}}{(1-u)\cos y+u\sin
y/y}\;\frac{p\!\!\!/_\perp}{m}
\Big]+c.t.\cr
\end{split}
\label{eq:mom1}
\end{equation}
Quite remarkably, in addition to the replacement $\sigma^3 y \to q\sigma^3
y$ in the exponentials, which originates from our taking the original
Tsai's formula for $G_0$
instead of that of \cite{DittrichReuter}, and to a global factor $1/2$,
it only differs from (3.27) of \cite{DittrichReuter}
 by $\displaystyle\int\frac{du}{\sqrt{1-u}}$
instead of $\displaystyle\int du$ and by the fact
that, in the present situation,
 $p_\parallel^2=-p_0^2,\ p\!\!\!/_\parallel = -\gamma_0 p_0$.
We thus see that, after these  lengthy but straightforward
transformations have been done, the electron self-energy for
QED$_{3+1}$
reduced on a 2-brane is formally very close to the one for QED$_{3+1}$.
The difference between the two integration measures for $u$ is however at
the origin of the completely different behaviors of the corresponding $\delta
m_{LLL}$ at the limit $m\to 0$, as we shall see in
subsection \ref{subsec:anaI}.

\subsection{Transforming the space representation of $\boldsymbol{\Sigma}$}

Unlike for the vacuum polarization in which the two opposite phases cancel, 
the phase $\Phi$, given in (\ref{eq:phase}), which occurs in the space
representation (\ref{eq:self0}) of the self-energy plays an important role.
This makes the calculations
all the more tedious as, like for QED$_{3+1}$, the integrations on $s$
and $u$ for $\Sigma(\hat p)$ obtained in (\ref{eq:mom1}) cannot be done
explicitly.

It is however
possible, along the lines of p.~47-52 of \cite{DittrichReuter} to get from
the space representation $\Sigma(x',x'')$ as  written in (\ref{eq:self0})
 a useful expression for $\Sigma(\hat \pi)$ 
defined by 
\begin{equation}
\Sigma(x',x'')
=<x'\ |\ \Sigma(\hat \pi)\ |\
x''>.
\label{eq:space1}
\end{equation}
$\Sigma(\hat\pi)$, which now depends on the covariant derivative $\hat\pi$,
has somewhat ``swallowed'' the phase $\Phi$, and
 is the  essential ingredient to  get
the self-mass $\delta m$ of an
electron on mass-shell ($\hat\pi\!\!\!/ +m=0$).

The $\int d^4p$ in  (\ref{eq:self0}), which is at the root of the
corresponding formal manipulations stays unchanged. 
One has to go through the  steps of p.~34-36 and p.~47-50
of \cite{DittrichReuter},
which use in particular eq.~(2.41) of \cite{DittrichReuter}
\begin{equation}
<X'\;|\; e^{-is\pi^2} \;|\; X''>=\Phi(X',X'') \int\frac{d^4
k}{(2\pi)^4}\;e^{ik(X'-X'')}\;\frac{1}{\cos q|e|Bs}\;e^{-is\big(k_\parallel^2
+k_\perp^2 \frac{\tan q|e|Bs}{q|e|Bs}\big)}
\end{equation}
and its avatars, (2.45) and more specially (2.46)
\begin{equation}
\begin{split}
& <X'\;|\; e^{-is\big(a_0 \pi_0\pi^0 + a_3 \pi_3 \pi^3 +a_\perp \pi_\perp^2
\big)}\;\Big(1,\gamma_0 \pi^0, \gamma_3 \pi^3, \gamma_\perp
\pi_\perp\Big)\;|\;X''>\cr
\hskip 2cm
&=\Phi(X',X'')\int\frac{d^4 k}{(2\pi)^4}\;e^{ik(X'-X'')}\;\frac{1}{\cos q|e|B
s a_\perp}\;
 e^{-is\big(a_0 k_0k^0 + a_3 k_3 k^3 +a_\perp \frac{\tan q|e|B s
a_\perp}{q|e|Bs a_\perp}\big)}\;\cr
&\hskip 4cm \Big(1, \gamma_0 k^0, \gamma_3 k^3, \frac{1}{\cos q|e|a_\perp B
s}\;e^{-iq|e|Bs a_\perp \sigma^3} \gamma_\perp p_\perp \Big).
\end{split}
\label{eq:DR246}
\end{equation}
They entail, by simple changes of variables ($\varphi$ is given in
(\ref{eq:chifi}))
\begin{equation}
\begin{split}
& \Phi(X',X'')\int \frac{d^4 p}{(2\pi)^4}\;e^{ip(X'-X'')}\;e^{-is\varphi}
= \cos \beta <X'\;|\;
e^{-isu(1-u)p_\parallel^2}\;e^{-i\frac{\beta}{q|e|B}\pi_\perp^2}\;|\;X''>,\cr
&\Phi(X',X'')\int \frac{d^4 p}{(2\pi)^4}\;e^{ip(X'-X'')}\;e^{-is\varphi}
 \Big(a\, \hat p\!\!\!/_\parallel + b\,p\!\!\!/_\perp\Big)\cr
& \hskip 2cm= \cos \beta <X'\;|\;
e^{-isu(1-u)p_\parallel^2}\;e^{-i\frac{\beta}{q|e|B}\pi_\perp^2}
\Big(a \hat p\!\!\!/_\parallel + b \cos \beta\; e^{iq\sigma^3 \beta}\;
p\!\!\!/_\perp\Big) \;|\;X''>
\end{split}
\label{eq:2rel}
\end{equation}
in which $\Delta(u,y)$ and  the angle $\beta$  have been introduced, which satisfy
\cite{DittrichReuter}
\begin{equation}
\begin{split}
& \sin\beta=\frac{(1-u)\sin y}{\Delta(u,y)^{1/2}},\quad
\cos\beta=\frac{(1-u)\cos y + u\sin y/y}{\Delta(u,y)^{1/2}},\cr
\Delta(u,y) &= (1-u)^2 +2u(1-u)\frac{\sin y\cos y}{y} +u^2\Big(\frac{\sin
y}{y}\Big)^2.
\end{split}
\label{eq:beta}
\end{equation}
After all terms inside (\ref{eq:mom1}) have been transformed via
(\ref{eq:2rel}), one gets the result
\begin{equation}
\begin{split}
\Sigma(\hat\pi) &=\frac{\alpha m}{4\pi}\int_0^\infty
\frac{ds}{s}\int_0^1\frac{du}{\sqrt{1-u}}\;e^{-isu^2m^2}\cr
& \hskip -1.5cm
 \Bigg[
\frac{e^{-is\Theta}}{\sqrt{\Delta(u,y)}} \Big[1+e^{-2iqy\sigma^3}
+(1-u)e^{-2iqy\sigma^3}\;\frac{\hat\pi\!\!\!/}{m}
 +(1-u)\Big(\frac{1-u}{\Delta(u,y)}+\frac{u}{\Delta(u,y)}\frac{\sin
y}{y}\;e^{-iqy\sigma^3}-e^{-2iqy\sigma^3}\Big)\frac{\pi\!\!\!/_\perp}{m}
\Big]+c.t. \Bigg],\cr
\Theta &=
u(1-u)(m^2-\hat\pi\!\!\!/^2)+\frac{u}{y}\big(\beta-(1-u)y\big)\pi_\perp^2
-u^2 \frac{|e|q}{2}\,\sigma_{\mu\nu}F^{\mu\nu},\cr
\end{split}
\label{eq:space1b}
\end{equation}
which differs from (3.38a) of \cite{DittrichReuter} by the absence of
$\gamma_3 \pi^3$ from $\hat \pi \!\!\!/$, the same factor $\frac12$ that we
already mentioned concerning (\ref{eq:mom1}),
 and the presence of $q\equiv -1$ in the exponentials (that was omitted in
\cite{DittrichReuter}).

\subsubsection{Renormalization conditions and counterterms}

The electron mass we define as the pole of its propagator, which is the
only gauge invariant definition.

We briefly recall here the general procedure to fix the counterterms.
 It is then straightforwardly adapted to
our concern by replacing everywhere $p$ with $\hat p$ and $\pi$ with $\hat
\pi$ ($\pi_\mu = p_\mu +|e|A_\mu$).

At $B=0$, the renormalized electron mass is defined by
\begin{equation}
m=m_0+\delta m,\quad \delta m= \Sigma(p)\big|_{p\!\!\!/+m=0},
\label{eq:mdef1}
\end{equation}
in which $m_0$ is the bare mass and $\Sigma(p)$ the bare self-energy.

In the presence of and external field $A^\mu$, the propagator of a Dirac
electron is
\begin{equation}
iG = \frac{i}{\pi\!\!\!/ +m_0 +\Sigma(\pi)},
\end{equation}
and we define, in analogy with (\ref{eq:mdef1}) the mass of the electron as
the pole of its propagator by
\begin{equation}
m=m_0 +\Sigma(\pi)\big|_{\pi\!\!\!/+m=0},\quad \delta m=
\Sigma(\pi)\big|_{\pi\!\!\!/+m=0}.
\label{eq:mdef2}
\end{equation}
$\delta m$ depends on the external $B$.

The on mass-shell renormalization conditions write 
\footnote{They are carefully explained p.38-41 of \cite{DittrichReuter}.}
\begin{equation}
\lim_{\pi\!\!\!/+m=0} \lim_{B\to 0} \Sigma^{ren}(\pi)=0,\quad
\lim_{\pi\!\!\!/+m=0} \lim_{B\to 0} \frac{\partial
\Sigma^{ren}(\pi)}{\partial \pi\!\!\!/}=0,
\label{eq:rencond}
\end{equation}
in which the superscript ``$ren$'' denotes the renormalized quantities.

They lead to the same counterterms as in \cite{DittrichReuter} but for the
simple modifications $p\to\hat p, \pi \to \hat\pi$, and one gets
\begin{equation}
\begin{split}
\Sigma(\hat\pi) &=\frac{\alpha m}{4\pi}\int_0^\infty
\frac{ds}{s}\int_0^1\frac{du}{\sqrt{1-u}}\;e^{-isu^2m^2}\cr
& \Bigg[
\frac{e^{-is\Theta}}{\sqrt{\Delta}} \Big[1+e^{-2iqy\sigma^3}
+(1-u)e^{-2iqy\sigma^3}\;\frac{\hat\pi\!\!\!/}{m}
 +(1-u)\Big(\frac{1-u}{\Delta}+\frac{u}{\Delta}\frac{\sin
y}{y}\;e^{-iqy\sigma^3}-e^{-2iqy\sigma^3}\Big)\frac{\pi\!\!\!/_\perp}{m}
\Big]\cr
& \hskip 3cm
\underbrace{-(1+u) -(m+\hat\pi
\!\!\!/)\Big(\frac{1-u}{m}-2imu(1-u^2)s\Big)}_{c.t.} \Bigg],
\end{split}
\label{eq:space2}
\end{equation}
in which $y, \Theta, \Delta$ are given in (\ref{eq:space1b}) and
(\ref{eq:beta}).

The 2nd counterterm vanishes on mass-shell (since it must satisfy the 1st
renormalization condition), and can therefore be forgotten in the
calculation of $\delta m$.

\subsection{The self-mass $\boldsymbol{\delta m_{LLL}}$ for an electron in
 the lowest Landau level}

The spectrum of a Dirac electron in a pure magnetic field directed along
$z$ is \cite{BerLifPit}
\begin{equation}
\epsilon_n^2 = m^2 + p_z^2 +(2n+1+\sigma_z)\,|e|B,
\end{equation}
in which $\sigma_z=\pm 1$ is $2\; \times$ the spin projection of the
electron
on the $z$ axis.
So, at $n=0, \sigma_z=-1, p_z=0$, $\epsilon_n=m$: this 
is  the lowest Landau level.

We can consider $A_\mu=\left(\begin{array}{c} A_0=0\cr A_x=0\cr A_y=xB\cr
A_z=0\end{array}\right)$ such that  $F_{12}=B$ is the only
non-vanishing component of the classical external $F_{\mu\nu}$.
Then, the wave function of the LLL writes
\cite{Luttinger} \cite{KuznetsovMikheev}
\begin{equation}
\psi_{n=0,s=-1,p_y=p_z=0}=\frac{1}{\sqrt{N}}
\left(\frac{|e|B}{\pi}\right)^{1/4}\;e^{-\frac{|e|B}{2}x^2}\;
\left(\begin{array}{c} 0\cr 1 \cr 0 \cr 0\end{array}\right),
\ N\stackrel{\cite{KuznetsovMikheev}}{=} \underbrace{L_y\;
L_z}_{dimensions\
along\ y\ and\ z}.
\label{eq:LLL}
\end{equation}

Following (\ref{eq:mdef2}), in order to determine  $\delta m$ for the (on
mass-shell) LLL, we shall sandwich the general self-energy operator
(\ref{eq:space2}) between two states $|\;\psi>$ defined in (\ref{eq:LLL})
and satisfying $(\hat\pi\!\!\!/ +m)|\;\psi>=0$.

The expression (\ref{eq:space2}) involves
$\hat\pi\!\!\!/$ that we shall replace by $-m$, $\Delta$
that needs not be transformed, and $\Theta$  which involves
$m^2-\hat\pi\!\!\!/^2$, $\pi_\perp^2$ and $\sigma_{\mu\nu}F^{\mu\nu}$.
The only non-vanishing component of $F^{\mu\nu}$ being $F^{12}=B$,
$\sigma_{\mu\nu}F^{\mu\nu}= \sigma_{12}F^{12}+\sigma_{21}F^{21}=
2\sigma_{12}F^{12}\equiv 2\sigma_3 B$.

Since the electron is an eigenstate of the Dirac equation in the presence
of $B$, $m^2 -\hat\pi\!\!\!/^2$ can be taken to vanish.
$\pi_\perp^2 \equiv \pi_1^2 + \pi_2^2$ is also identical, since the
LLL has $p_z=0$ and we work in a gauge with $A_z=0$, to $\vec
\pi^2\equiv\pi^2 + \pi_0^2$. One has $\pi\!\!\!/^2 = -\pi^2
+\frac{q|e|}{2}\sigma_{\mu\nu}F^{\mu\nu}$ such that $\pi_\perp^2 =
-\hat\pi\!\!\!/^2 +\pi_0^2 +\sigma_3\,q|e|B$. Since our gauge for the external $B$
has $A_0=0$, $\pi_0^2 =p_0^2$, which is the energy squared of the electron,
identical to $m^2$ for the LLL. Therefore, on mass-shell,
$\pi_\perp^2 = \sigma_3\,q|e|B$.
 When sandwiched between LLL, \newline
 $<\psi\;|\;\sigma^3\;|\;\psi>=
\left(\begin{array}{cccc}0 & 1 & 0 & 0 \end{array}\right)
 diag(1,-1,1,-1)\left(\begin{array}{c} 0 \cr 1 \cr 0 \cr
0\end{array}\right) = -1$
such that  $\sigma^3$ can be replaced by $(-1)$.
$\Theta$ shrinks to $u(\beta/y-1)q|e|B \sigma^3$, which gives, replacing
$\sigma^3$ with $(-1)$, $\Theta \to u(1-\beta/y)q|e|B$.
$\sigma^3$ can also be replaced by $(-1)$ in the exponentials of
(\ref{eq:space2}).

$\Sigma(\hat\pi)$ in (\ref{eq:space2}) also involves a term proportional to
 $\pi\!\!\!/_\perp$. Since the LLL
has $p_z=0$ and we work at $A_z=0$, this is also equal to
$ \vec\gamma.\vec{ \hat\pi} = \gamma^\mu \hat\pi_\mu -\gamma^0 \pi_0
= \hat\pi\!\!\!/ +\gamma^0 p^0$. $<\psi\;|\;\hat\pi\!\!\!/\;|\; \psi>=-m$ such
that\newline
$<\psi\;|\; \pi\!\!\!/_\perp\;|\;\psi> = <\psi\;|\; -m+\gamma^0
p^0\;|\;\psi>$.
Since $\gamma^0=diag(1, 1, -1, -1)$,
eq.~(\ref{eq:LLL})  yields $<\psi\;|\; \pi\!\!\!/_\perp\;|\;\psi>
=-m+p^0$.
The energy $p^0$ of the LLL $|\;\psi>$ being equal to $m$,
this term vanishes.

Gathering all information and simplifications leads finally to
\begin{equation}
\delta m_{LLL} \equiv \Sigma(\hat \pi)_{\hat\pi\!\!\!/ +m=0}=
\frac{\alpha m}{4\pi}\int_0^\infty \frac{ds}{s}\int_0^1
\frac{du}{\sqrt{1-u}}\;e^{-isu^2m^2}
\Bigg[\frac{e^{-is\Theta(u,y)}}{\sqrt{\Delta(u,y)}}\;(1+u\,e^{2iqy})
-\underbrace{(1+u)}_{from\ c.t.}\Bigg],
\label{eq:dmLLL}
\end{equation}
in which $y=|e|Bsu$ as before, $\Delta(u,y)$ is  the same
as in (\ref{eq:space1b}), $\beta$ the same as in (\ref{eq:beta}), and
$\Theta$ has shrunk down to
\begin{equation}
\Theta(u,y)
=uq|e|B\Big(1-\frac{\beta(u,y)}{y}\Big)=uq|e|B-\frac{q\beta(u,y)}{s}.
\end{equation}

\section{The ``reduced'' Demeur-Jancovici integral $\boldsymbol{\hat I(L)}$}

\subsection{General expression}

We define $\hat I(L)$ by
\begin{equation}
\delta m_{LLL} = \frac{\alpha\; m}{4\pi}\; \hat I(L)\quad with\
L=\frac{|e|B}{m^2}.
\label{eq:DJdef}
\end{equation}
such that
\begin{equation}
\hat I(L) =\int_0^\infty \frac{ds}{s}\int_0^1
\frac{du}{\sqrt{1-u}}\;e^{-isu^2m^2}
\Bigg[\frac{e^{-is\Theta(u,y)}}{\sqrt{\Delta(u,y)}}\;(1+u\,e^{2iqy})
-\underbrace{(1+u)}_{from\
c.t.}\Bigg] 
\label{eq:I0}
\end{equation}
By a successive change of variables, we cast it in a form similar to
$I(L)$ deduced by Jancovici in \cite{Jancovici} from the formula obtained
by Demeur in \cite{Demeur}, and that was revisited in \cite{Machet1}.
The calculations, which are detailed in appendix
\ref{section:DemeurJanco}, lead to

\begin{equation}
\hat I(L)= \int_0^{\infty} dz
\int_0^1 \frac{dv}{\sqrt{1-v}}\; e^{-z\frac{m^2}{|e|B}}
\Bigg[ \frac{2\left(1+v\,e^{-2z/v}\right)}{2z(1-v)
+v^2\left(1-e^{-2z/v}\right)}
-\underbrace{\frac{1+v}{z}}_{from\ c.t.} \Bigg],
\label{eq:I2}
\end{equation}
which is the expression which we shall focus on hereafter.

Calling
\begin{equation}
f(v,z)=\frac{2(1+ve^{-2z/v})}{2z(1-v)+v^2(1-e^{-2z/v})}-\frac{1+v}{z},
\end{equation}
$\hat I(L)$ in (\ref{eq:I2}) can be cast in the form
\begin{equation}
\hat I(L)= \int_0^{\infty}dz\;e^{-z/L}\int_0^1
\frac{dv}{\sqrt{1-v}}\;f(v,z).
\end{equation}

That $\hat I(L)$ would be divergent at $z=0$ without the counterterm can be
easily seen by expanding $ \frac{2\left(1+v\,e^{-2z/v}\right)}{2z(1-v)
+v^2\left(1-e^{-2z/v}\right)} \stackrel{z\to 0}{\sim} \frac{1+v}{z} +v-1
+{\cal O}(z)$

\subsection{Analytical evaluation of $\boldsymbol{\hat I(L)}$}
\label{subsec:anaI}



We split $\int_0^\infty dz\;(...)$ in $\hat I(L)$ given by (\ref{eq:I2}) into
$\int_0^{z_0} dz\;(...)+ \int_{z_0}^\infty dz\;(...)$, with:\newline
*\  $z_0$ large enough such that, in the 2nd integral, in which $z>z_0$,
$f(v,z) \simeq \displaystyle\frac{2}{2z(1-v)+v^2}-\displaystyle\frac{1+v}{z}$, that is, the
exponentials can be neglected;\newline
*\ $z_0$ small enough for $\int_0^{z_0} dz\;(...) \ll \int_0^\infty dz\;(...)
\simeq \int_{z_0}^\infty dz\;(...)$ and can be neglected.

In practice, $z_0=1$ fits perfectly and, even down to $L=20$, the ratio of
the
2 integrals is $\leq 1/100$.

$\int_0^{z_0} dz\;(...)$ involves two
canceling divergent integrals, and,  for proper numerical evaluation,
one has to set the lower bound of integration to $\epsilon \not=0$,
checking stability when $\epsilon$ decreases from $10^{-3}$ down to
$10^{-12} \ldots$.

Likewise, to numerically evaluate $\int_{z_0\simeq 1}^\infty (\ldots)$, avoiding to deal
with too small numbers requires to set the upper bound of integration at a
large but finite number (which depends on the value of $L$)
 instead of infinity and to check stability by
varying this bound inside a large interval. 

The result is that, for $L\geq 20$ and $z_0\simeq 1$ one can approximate at a precision
better than $1/100$
\begin{equation}
\hat I(L) \approx \int_{z_0\simeq 1}^\infty
dz\;e^{-z/L}\int_0^1\frac{dv}{\sqrt{1-v}}\;\Big[
\frac{2}{2z(1-v)+v^2}-\frac{1+v}{z}\Big].
\label{eq:Iapp2}
\end{equation}
One has 
\begin{equation}
\begin{split}
g(z) &\equiv \int_0^1\frac{dv}{\sqrt{1-v}}\;\frac{2}{2z(1-v)+v^2}=
\frac{2}{\sqrt{z(z-2)}}
\left[
\frac{\tan^{-1}\displaystyle\frac{1}{\sqrt{-1+z-\sqrt{z(z-2)}}}}{\sqrt{-1+z-\sqrt{z(z-2)}}}
-\frac{\tan^{-1}
\displaystyle\frac{1}{\sqrt{-1+z+\sqrt{z(z-2)}}}}{\sqrt{-1+z+\sqrt{z(z-2)}}}\right],\cr
& \int_0^1 dv\;\frac{1+v}{\sqrt{1-v}} = \frac{10}{3},\cr
& \int_{z_0\simeq 1}^\infty dz\;\frac{e^{-z/L}}{z}=\Gamma(0,1/L),
\end{split}
\label{eq:g}
\end{equation}
therefore
\begin{equation}
\delta m_{LLL}= \frac{\alpha m}{4\pi} \Big(\int_{z_0 \approx 1}^\infty
dz\;e^{-z/L}\;g(z) -\frac{10}{3}\Gamma(0,1/L)\Big).
\label{eq:delmg}
\end{equation}
On Fig.~2 we compare $g(z)$ given in (\ref{eq:g}) (blue)
 with the one  obtained in \cite{Machet1} for standard QED$_{3+1}$
$\big(g(z)=\ln(z-1+\sqrt{z(z-2)})/\sqrt{z(z-2)} \simeq \ln z/z
+\pi/2z^{1.175}\big)$ (yellow).
\begin{center}
\includegraphics[width=6cm, height=4cm]{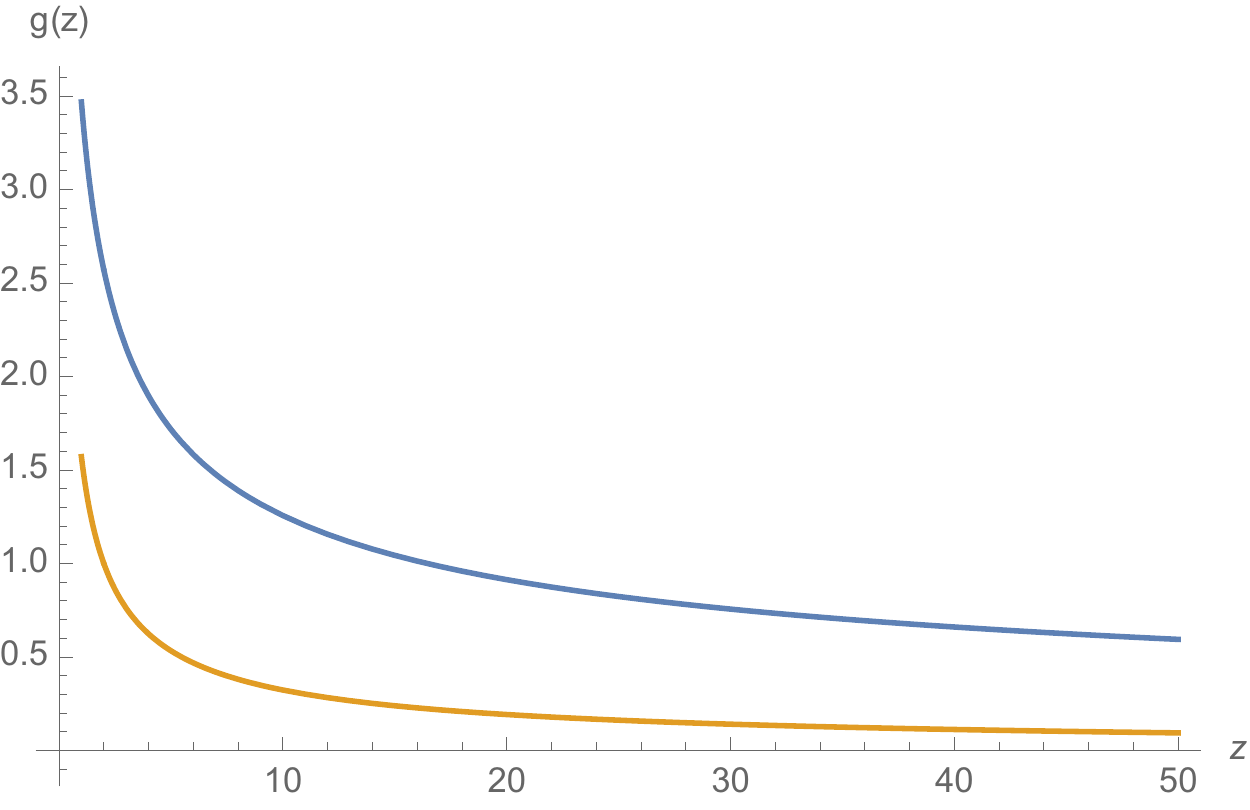}

\smallskip
{\em Fig.~2: A comparison between the integrand $g(z)$ in QED$_{3+1}$
(yellow) and in QED$_{3+1}$ reduced on a 2-brane (blue)}
\end{center}

We now proceed like M.I.~Vysotsky in \cite{Vysotsky2010} and
 look for an interpolating function for $g(z)$. One has
\begin{equation}
\begin{split}
g(1) &\approx 3.468,\cr
g(z) &\stackrel{z\to\infty}\simeq \pi\sqrt{\frac{2}{z}}-\frac{2}{z}+{\cal
O}(\frac{1}{z^{3/2}}) \simeq \frac{4.443}{\sqrt{z}} + \ldots
\end{split}
\end{equation}
and an excellent fit for $z\in[z_0\simeq 1,\infty]$ is
\begin{equation}
g(z) \approx \pi\sqrt{\frac{2}{z}}+\frac{g(1)-\pi\sqrt{2}}{z}.
\label{eq:gapp}
\end{equation}
It is plotted in yellow on Fig.~3,  while the exact $g$ is in blue.

\begin{center}
\includegraphics[width=6cm, height=4cm]{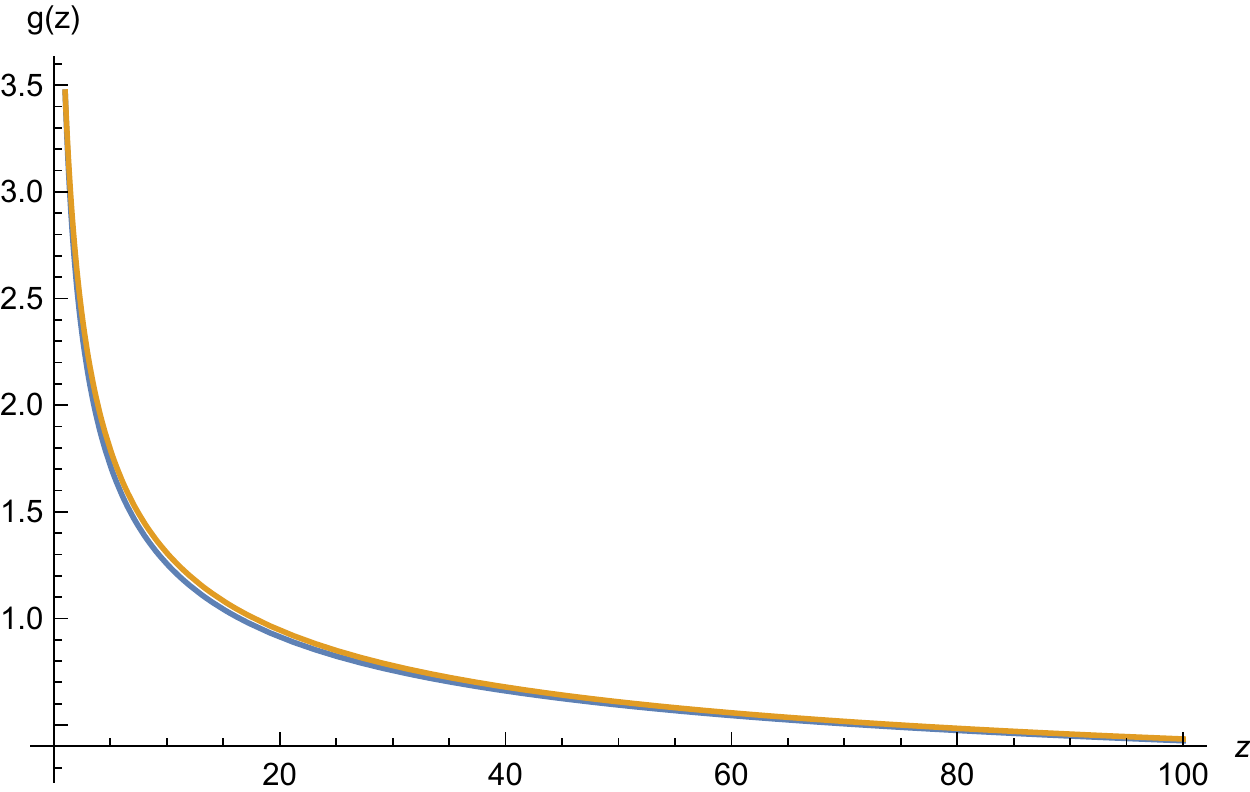}

\smallskip
{\em Fig.~3: exact (blue) and approximated (yellow) $g(z)$ for $z \geq 1$}
\end{center}

This approximation gives (using (\ref{eq:DJdef}))\footnote{$Erfc(x) = 1-Erf(x)$.}
\begin{equation}
\delta m_{LLL} \equiv \frac{\alpha m}{4\pi}\;\hat I(L)
=\frac{\alpha}{4\pi}\sqrt{|e|B}\Bigg[
\sqrt{2}\;\pi^{3/2}\;Erfc\big(\frac{1}{\sqrt{L}}\big)
+\frac{\Gamma(0,\frac{1}{L})}{\sqrt{L}} \Big(
g(1)-\pi\sqrt{2}-\frac{10}{3}\Big)\Bigg].
\label{eq:dm3}
\end{equation}
When $L\to\infty$, $Erfc(\frac{1}{\sqrt{L}}) \simeq
1-\frac{2}{\sqrt{\pi}}\frac{1}{\sqrt{L}} + \ldots$ and
$\Gamma(0,\frac{1}{L}) \simeq \ln L-\gamma_E +\ldots$ such that
\begin{equation}
\delta m_{LLL} \stackrel{L\to\infty}{\simeq}
\frac{\alpha}{2}\;\sqrt{|e|B}\,\sqrt{\frac{\pi}{2}}\Bigg[
1-\frac{2}{\sqrt{\pi L}} +\frac{1}{\sqrt{2}\;\pi^{3/2}} \frac{\ln
L-\gamma_E}{\sqrt{L}}
\big(g(1)-\pi\sqrt{2}-\frac{10}{3}\Big)+\ldots\Bigg].
\label{eq:dmlim}
\end{equation}
The constant term comes from the contribution to $\hat I(L)$  of
 $\displaystyle\int_{z_0\simeq 1}^\infty dz\; e^{-z/L}/\sqrt{z}
= \sqrt{\pi L}\; Erf(\sqrt{z/L})\Big|_{z_0\simeq 1}^\infty$ at $\infty$. So, it is not
sensitive to the precise value of $z_0=1$, but it is controlled by the leading
behavior of $g(z)\sim 1/\sqrt{z}$ at $z\to\infty$
\footnote{By comparison, in the case of standard QED$_{3+1}$, the leading
behavior of $g(z)$ when $z\to\infty$ being
$g(z)\stackrel{z\to\infty}{\sim}\ln z/z$, one gets $I(L) \sim
\int_{z_0\simeq 1}^{\infty} dz\, e^{-z/L}\ln z/z  \sim constant$, which yields
$\delta m_{LLL} \sim \frac{\alpha m}{2\pi}.constant \stackrel{m\to 0}{\to} 0$.}.

It is important to check that, at the limit of large $L$, the first
integral $\int_0^{z_0=1} dz(\ldots)$ is stable and can still be neglected with
respect to the second integral. This is shown on Fig.~4-left, in which
we plot the 1st integral as a function of $L$. As already mentioned, the
numerical cancellation of infinities requires that the lower bound of integration be
set not to $0$ but to smaller and smaller $\epsilon$. The curve in blue
corresponds to $\epsilon=10^{-3}$, and the 3 other curves, green, yellow
and red, corresponding to $\epsilon=10^{-6}, 10^{-9}, 10^{-12}$ are
superposed; $\hat I(L)$ as given by (\ref{eq:dm3}) is plotted on
Fig.~4-right. We see that, even at very large values of $L$, the 1st integral
can always be safely neglected inside $\hat I(L)$.

\vbox{
\begin{center}
\includegraphics[width=6cm, height=4cm]{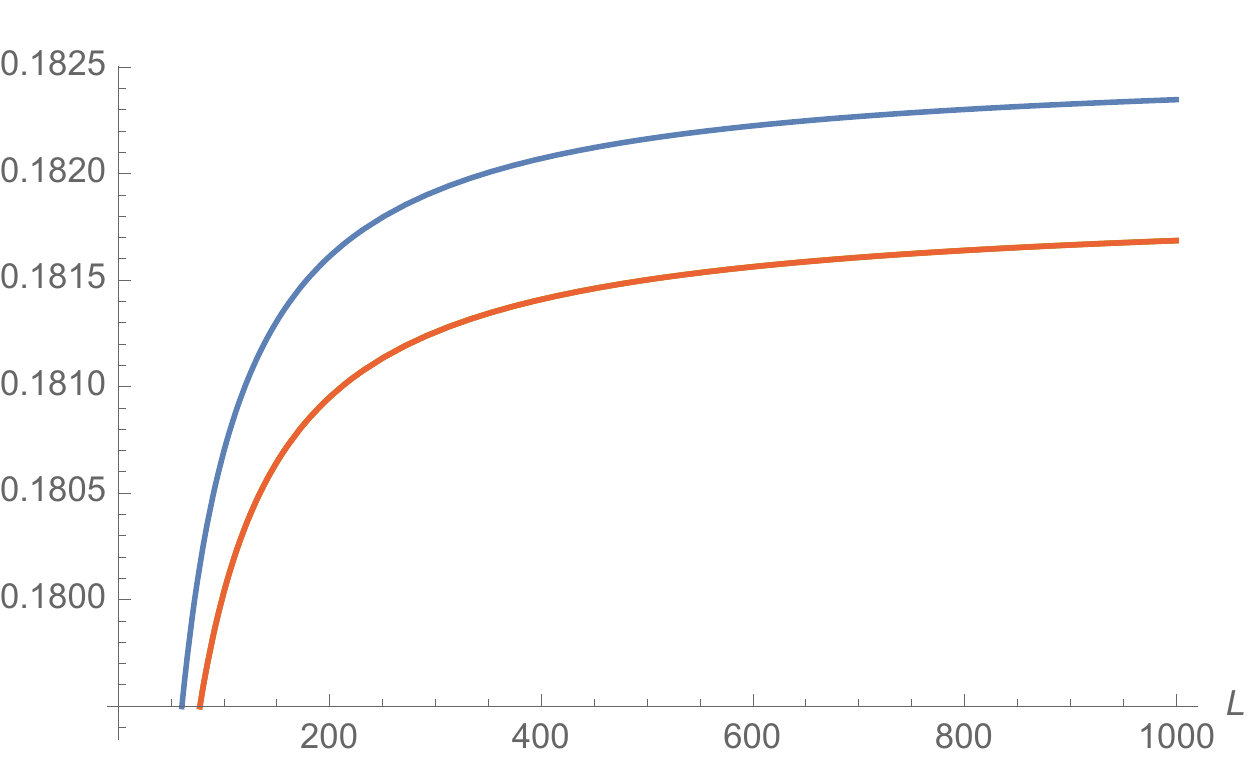}
\hskip 2cm
\includegraphics[width=6cm, height=4cm]{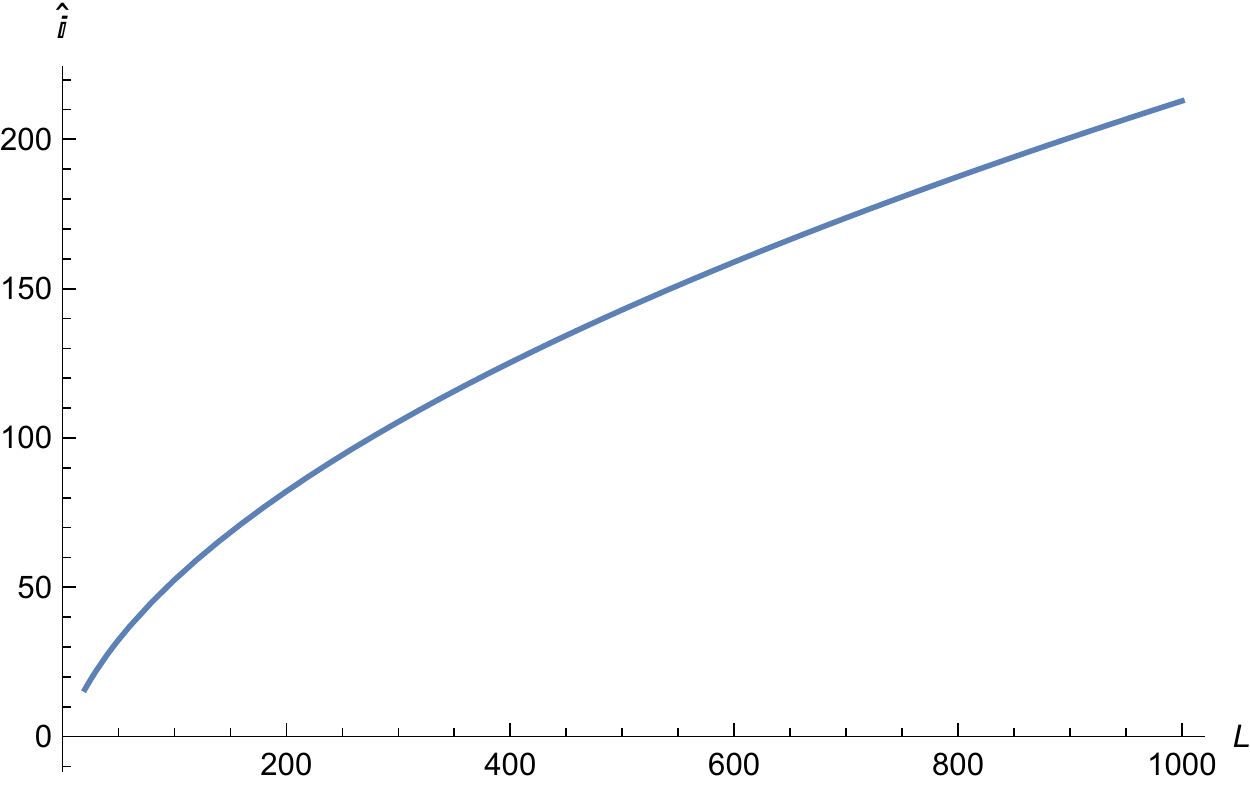}

\smallskip
{\em Fig.~4: on the left: value of the 1st (neglected inside $\hat I(L)$) integral
$\int_0^{z_0=1}dz(...)$ for lower bounds of integrations going from $10^{-3}$
(blue) to $10^{-12}$ (yellow, green, red); on the right : $\hat I(L)$ }
\end{center}
}

\section{A non-vanishing 1-loop $\boldsymbol{\delta m_{LLL}}$
 at $\boldsymbol{m\to 0}$
}

From (\ref{eq:dmlim}) one gets immediately (restoring $\hbar$ and $c$)
\begin{equation}
\delta m_{LLL} \stackrel{m\to 0}{\to}
\frac{\alpha}{2}\;\sqrt{\frac{\pi}{2}}\;\sqrt{\frac{\hbar |e|B}{c^2}},
\label{eq:dm4}
\end{equation}
which shows that, in an external magnetic field, this model, equivalent to
reduced QED$_{3+1}$ on a 2-brane, cannot stay massless at 1-loop.
Notice that (\ref{eq:dm4}) fulfills the renormalization conditions
(\ref{eq:rencond}), which are expressed at  $B=0$.

Since the role of the counterterms is slightly more subtle than for QED$_{3+1}$
(in which they yield the large logs (see \cite{Machet1})),
 it is useful to make some comments about them.

In (\ref{eq:dmLLL}), the (infinite) counterterm only depends on $m$ through
the exponential  $e^{-isu^2m^2}$ inside the integrand.

Noting respectively $b.term$ and $c.term$ the bare term and the counterterm
inside the expression (\ref{eq:dmLLL}) of $\delta m_{LLL}$,
one can write symbolically
$b.term= +\infty + f_1(m, eB), \quad 
c.term = -\infty + f_2(m)$, in which $f_1, f_2$ are finite.\newline
The change of variables (\ref{eq:chvar1}) introduces a dependence of both on
$L$, that we write symbolically
$ b.term= +\infty +\hat f_1(m,eB, L) + \zeta(L)=+\infty +
h_1(m,eB,L), \quad c.term= -\infty +\hat f_2(m,L) -\zeta(L)=-\infty +
h_2(m,L)$.
Therefore, via the change of variable (\ref{eq:chvar1}),
the counterterm has reacted on the bare contribution and the two become 
entangled (we introduced $\pm \zeta$ to picture the fact that this dependence
globally cancels but, in practice, one cannot ``isolate'' $\zeta$ ; also,
strictly speaking, these terms are not defined before the infinities are
regularized and canceled).

The ``educated'' splitting \cite{Jancovici} of the $z$ interval of integration
$[0,\infty[= [0,z_0]+[z_0,\infty[$ brings then
$\hat I(L)$ down to the approximation (\ref{eq:Iapp2}).
Let us call the integrands in there $h_1^{z_0}(m, eB, L)$ and $h_2^{z_0}(m,L)$.
That the limit $m\to 0$ yields a constant $\delta m$, or,
equivalently, $\hat I(L) \propto \sqrt{L}$.
is due to $h_1^{z_0}$ (and the corresponding $g(z)$ defined in (\ref{eq:g})
(\ref{eq:delmg})
(\ref{eq:gapp})) which  has an
asymptotic expansion $\simeq 1/\sqrt{z}$ at $z\to \infty$.
This makes the result insensitive to the precise value of $z_0$.
By contrast, as we have mentioned, in standard QED$_{3+1}$,
the asymptotic behavior of $g(z)$ is $\ln z/z$ \cite{Machet1}.

$h_1^{z_0}$ no longer represents the bare contribution for the reasons that
we just evoked: * a change of variables introduced an extra dependence on $L$
that mixes with the counterterm; * the splitting of the $z$ interval
of integration collects in the neglected (small)  $\int_0^{z_0} dz(\ldots)$,
in particular, the two canceling infinite parts of the bare term and of the
counterterm, establishing a second connection between the two. In this
respect, both play crucial roles in the massless limit of $\delta_{LLL}$,
that can hardly be disentangled.

Last, let us remark that it is  necessary to make the $z$  integration at
$m\not=0$ {\em before taking the limit $m\to 0$}, otherwise, since
$L=|e|B/m^2$, one gets the undetermined expression $\displaystyle\frac{1}{0}
\times 0$. Had we started from a massless theory, we would have obtained
such an undetermined result. This is why one can only state that {\em the
massless limit}
of the 1-loop $\delta m_{LLL}$ goes to a constant, or, equivalently, that the
model under consideration cannot stay massless at 1-loop.

\section{Restricting to the lowest Landau level of the virtual electron}

\subsection{Basics}

The contribution of different Landau levels to the propagator of an
electron in a constant uniform external $B$ has been investigated in
\cite{KuznetsovOkrugin2011} and \cite{ChodosEverdingOwen1990}.
From eqs.(22,23,24) of \cite{KuznetsovOkrugin2011} one gets
\begin{equation}
\begin{split}
G(x,x') &= \sum_{n=0}^{\infty} G^n(x,x')
=\sum_{n=0}^{\infty}e^{i\omega(x,x')}\;\hat G^n(x-x')\cr
&= e^{i\omega(x,x')} \int
\frac{d^4p}{(2\pi)^4}\;e^{-ip(x-x')}\sum_{n=0}^{\infty}\hat G^n(p,B),\cr
\omega(x,x') &= -\frac{|e|B}{2}(x_1+x'_1)(x_2-x'_2),
\end{split}
\label{eq:basics}
\end{equation}
in which $x=(x_0,x_1,x_2,x_3), x'=(x'_0,x'_1,x'_2,x'_3)$.
The factor $e^{i\omega(x,x')}$ is identical to Schwinger's $\Phi(x,x')$
 as written in (\ref{eq:phase}) (see for example \cite{KuznetsovMikheev},
chapter 3).

Using the conventions and metric $(-+++)$ of Schwinger,
the contribution of the LLL is
\begin{equation}
-i\hat G^{n=0}(p,B)= e^{-p_\perp^2/|e|B}\int_0^\infty ds_1\;
e^{-is_1(m^2+p_\parallel^2)}(m-p\!\!\!/_\parallel)(1-i\gamma_1\gamma_2),
\label{eq:GLLL}
\end{equation}
in which we have introduced the Schwinger's parameter $s_1$ (see also
appendix \ref{section:LLLprop}).

To determine the contribution of the LLL of the virtual electron to the
self-energy, we have to calculate (see (\ref{eq:brane}))
\begin{equation}
i\Sigma^{n=0}(\hat p,B)=-\frac{e^2}{2}\int\frac{d^3\hat p}{(2\pi)^3}\;
\gamma^\mu \hat G^{n=0}(\hat p-\hat k,B)\frac{g_{\mu\nu}}{\sqrt{\hat
k^2}}\gamma^\nu
\end{equation}
One introduces as before (see (\ref{eq:trick}))
 the Schwinger parameter $s_2$ for the photon propagator
and, instead of eq.~(3.11) of \cite{DittrichReuter}, one gets
\begin{equation}
\hskip -1cm
\Sigma^{n=0}(p,B) = -ie^2\int_0^\infty ds_1 \sqrt{\frac{i}{\pi}}\int_0^\infty
\frac{ds_2}{\sqrt{s_2}} \int \frac{d^3 \hat k}{(2\pi)^3}\;
e^{-\frac{(p-k)_\perp^2}{|e|B}}\;e^{-is_2(\hat k^2-i\epsilon)}\;
e^{-is_1\big(m^2+(\hat p-\hat k)_\parallel^2\big)}\;
\gamma^\mu
\big(m-(p\!\!\!/_\parallel-k\!\!\!/_\parallel)\big)(1-i\gamma_1\gamma_2)\gamma_\mu.
\end{equation}
We use again the change of variables (\ref{eq:changevars}) 
together with
\begin{equation}
z=|e|Bs_1,\quad y=|e|Bus.
\end{equation}
Like before, aiming at performing the integration $\int d^3\hat k$,
one rewrites the exponentials
(watch the ``$i$'' which now occurs). Since $s$ cannot be factorized
everywhere, we have now included it into the definitions of $\chi_0$ and
$\varphi_0$, unlike previously for $\chi$ and $\varphi$.
\begin{equation}
\begin{split}
& \frac{(p-k)_\perp^2}{i|e|B} +s_2 \hat k^2 +s_1\big (m^2+(\hat p-\hat
k)_\parallel^2\big)\cr
& =usm^2 +su(1-u)\hat p_\parallel^2 +s(\hat k -u \hat p)_\parallel^2
+\Big(s(1-u)+\frac{1}{i|e|B}\Big)\Big(k_\perp-\frac{p_\perp}{1+i|e|Bs(1-u)}\Big)^2
+p_\perp^2 \frac{s(1-u)}{1+i|e|Bs(1-u)}\cr
& =\chi_0 + \varphi_0,\cr
& \chi_0= s(\hat k -u \hat p)_\parallel^2
+\Big(s(1-u)+\frac{1}{i|e|B}\Big)\Big(k_\perp-\frac{p_\perp}{1+i|e|Bs(1-u)}\Big)^2,\cr
& \varphi_0 = usm^2 +su(1-u)\hat p_\parallel^2+p_\perp^2
\frac{s(1-u)}{1+i|e|Bs(1-u)}
= usm^2 +b_0 \hat p_\parallel^2 + b_\perp p_\perp^2,\cr
& \hskip 3cm b_0=us(1-u),\quad  b_\perp=\frac{s(1-u)}{1+i|e|Bs(1-u)},
\end{split}
\label{eq:chifi2}
\end{equation}
such that
\begin{equation}
\Sigma^{n=0}(p,B) = -i\frac{e^2}{2}\int_0^\infty ds\;\sqrt{s} \sqrt{\frac{i}{\pi}}\int_0^1
\frac{du}{\sqrt{1-u}} \int \frac{d^3 \hat k}{(2\pi)^3}\;
e^{-i(\chi_0+\varphi_0)}\;\gamma^\mu
\big(m-(p\!\!\!/_\parallel-k\!\!\!/_\parallel)\big)(1-i\gamma_1\gamma_2)\gamma_\mu.
\end{equation}
One then shifts the variables $k_\parallel \to r_\parallel= k_\parallel -u
p_\parallel$,
$k_\perp \to r_\perp= k_\perp-\frac{p_\perp}{1+i|e|Bs(1-u)}$. One has
$\chi_0=s \hat r_\parallel^2 +\Big(s(1-u)+\frac{1}{i|e|B}\Big)r_\perp^2$.

Then, $(m-\gamma^0(k^0-p^0))=m-\gamma^0(r^0+(u-1)p^0)$.
$\chi_0$ being even since it depends  on $r_0^2$,  the odd term $\propto r_0$
yields a vanishing contribution to the $\int dk_0$. One can thus replace
$m-\gamma^0(k^0-p^0)$ by  $m-(u-1)\gamma^0 p^0$. One gets
\begin{equation}
\hskip -1cm
\Sigma^{n=0}(p,B)= -i\frac{e^2}{2}\sqrt{\frac{i}{\pi}}\int_0^\infty ds\sqrt{s}
\int_0^1 \frac{du}{\sqrt{1-u}} \int \frac{d^3 \hat r}{(2\pi)^3}\;
e^{-i\varphi_0}\;e^{-i[sr_\parallel^2+(s(1-u)+1/i|e|B)r_\perp^2]}
\gamma^\mu\big(m+(u-1)
p\!\!\!/_\parallel\big)(1-i\gamma_1\gamma_2)\gamma_\mu.
\end{equation}
With the help of (\ref{eq:standint})
one gets
\begin{equation}
\int d^3 \hat r\;e^{-i\chi}=e^{-i\pi/4}\frac{\sqrt{\pi}}{\sqrt{s}}
(\sqrt{\pi}e^{-i\pi/4})^2\frac{1}{s(1-u)+1/i|e|B},
\end{equation}
and, since $\sqrt{i}=e^{i\pi/4}$,
\begin{equation}
\begin{split}
\Sigma^{n=0}(p,B) 
&=-\frac{e^2}{16\pi^2}\int_0^\infty ds \int_0^1
\frac{du}{\sqrt{1-u}}\;e^{-i\varphi_0}\;\frac{i|e|B}{1+i|e|B\,s(1-u)}
\gamma^\mu
\big(m+(1-u)\gamma^0 p^0\big)(1-i\gamma_1\gamma_2)\gamma_\mu.
\end{split}
\end{equation}
Next, one performs the Dirac algebra
\begin{equation}
\begin{split}
\gamma^\mu
\big(m+(1-u)\gamma^0 p^0\big)(1-i\gamma_1\gamma_2)\gamma_\mu
& = -4m +2im\gamma_1 \gamma_2 +2(1-u)p^0 \gamma^0 +2i(1-u) p^0 \gamma^0
\gamma_1 \gamma_2 \cr
&= -4m +2(1-u)p^0 \gamma^0 (1+i\gamma_1 \gamma_2),
\end{split}
\end{equation}
such that
\begin{equation}
\hskip -1cm
\Sigma^{n=0}(p,B) = -\frac{\alpha}{2\pi}\int_0^\infty ds\int_0^1
\frac{du}{\sqrt{1-u}}\;e^{-isum^2}e^{-i(b_0 \hat p_\parallel^2 + b_\perp
p_\perp^2)}\;\frac{i|e|B}{1+i|e|B\,s(1-u)}
\big[ -2m+(1-u)p^0 \gamma^0 (1+i\gamma_1 \gamma_2)\big]+c.t.,
\end{equation}
in which $b_0$ and $b_\perp$ are given in (\ref{eq:chifi2}) and where
 we have now mentioned the counterterms (c.t.) that need eventually
to be introduced to fulfill suitable renormalization conditions.

We are interested in $\delta m^0_{LLL}$ concerning external electrons in the
LLL. To get it we sandwich $\Sigma(\pi)$
between two LLL eigenstates. Since these are annihilated by
$1+i\gamma_1\gamma_2$, the only term that may play a role is the one
proportional to $m$. Accordingly, the quantity of interest to us is
\begin{equation}
\Sigma^{n=0}_{LLL}(p,B) = \frac{\alpha\, m}{\pi}\int_0^\infty ds\int_0^1
\frac{du}{\sqrt{1-u}}\;e^{\displaystyle -isum^2}e^{\displaystyle -is(1-u)
\Big(u\hat p_\parallel^2 + \frac{p_\perp^2}{1+i|e|Bs(1-u)}\Big)}
\;\frac{i|e|B}{1+i|e|B\,s(1-u)} +c.t.
\end{equation}

\subsection{Transforming the space representation}

One needs to determine $\Sigma(\pi)$ satisfying (\ref{eq:space1}).
To that purpose, one must find the suitable change of variables to adapt (2.45) (2.46) of
\cite{DittrichReuter} to the present situation, that is to determine $a_0$
and $a_\perp$ in (\ref{eq:DR246} (which is the same as (2.46)
 of \cite{DittrichReuter}).
 One must have
\begin{equation}
\begin{split}
& \exp[-is a_\perp p_\perp^2 \frac{\tan |e|Bs a_\perp}{|e|Bs a_\perp}]
= \exp[-is(1-u)\frac{p_\perp^2}{1+i|e|Bs(1-u)}]\cr
& \Leftrightarrow \tan |e|Bs a_\perp=\frac{|e|Bs(1-u)}{1+i|e|Bs(1-u)}
\Leftrightarrow a_\perp=\frac{1}{|e|Bs} \tan^{-1}
\frac{|e|Bs(1-u)}{1+i|e|Bs(1-u)}.
\end{split}
\end{equation}
Then
\begin{equation}
\cos |e|Bs a_\perp = \cos \tan^{-1} \frac{|e|Bs(1-u)}{1+i|e|Bs(1-u)}.
\end{equation}
One also has trivially
\begin{equation}
-is a_0(-p_0^2) =-isu(1-u)(-p_0^2) \Leftrightarrow a_0=u(1-u).
\end{equation}
This gives
\begin{equation}
\begin{split}
\Sigma^{n=0}(\pi_0,\pi_\perp) &= \frac{\alpha m}{\pi} \int_0^\infty ds \int_0^1
\frac{du}{\sqrt{1-u}}\;\Big[\cos \tan^{-1} \frac{|e|Bs(1-u)}{1+i|e|Bs(1-u)}\Big]
\frac{i|e|B}{1+i|e|Bs(1-u)}\cr
& \hskip -1cm
e^{\displaystyle -isum^2}\;e^{\displaystyle -isu(1-u)(-\pi_0^2)}\;
e^{\displaystyle -is \pi_\perp^2\bigg(\frac{1}{|e|Bs} \tan^{-1}
\frac{|e|Bs(1-u)}{1+i|e|Bs(1-u)}\bigg)}
+ c.t.
\label{eq:LLspace}
\end{split}
\end{equation}

\subsection{Renormalization conditions and counterterms}

Let us consider general on mass-shell external electrons.
Since renormalization conditions have to be expressed  at $B=0$, let us
also consider the limit $B\to 0$ of  $\Sigma^{n=0}(\pi)$.

\begin{equation}
\begin{split}
\Sigma^{n=0}(\pi_0,\pi_\perp) & \stackrel{B\to 0}{\sim} 
\frac{\alpha m}{\pi} \int_0^\infty ds \int_0^1
\frac{du}{\sqrt{1-u}}\;\Big[\cos 0\Big]
\frac{i|e|B}{1+0}\cr
& \hskip -1cm
e^{\displaystyle -isum^2}\;e^{\displaystyle -isu(1-u)(-\pi_0^2)}\;
e^{\displaystyle -is \pi_\perp^2\bigg(\frac{1}{|e|Bs} \arctan
\frac{|e|Bs(1-u)}{1+0}\bigg)}
+\ terms\ \propto\ (1+i\gamma_1\gamma_2) + c.t \cr
& \hskip -2cm
 \sim \frac{\alpha m}{\pi} \int_0^\infty ds
\int_0^1\frac{du}{\sqrt{1-u}}\;i|e|B\;
 e^{\displaystyle -isum^2}\;e^{\displaystyle isu(1-u)\pi_0^2}\;
e^{\displaystyle -i \pi_\perp^2\;\frac{1}{|e|B} |e|Bs(1-u)}
+\ terms\ \propto\ (1+i\gamma_1\gamma_2) + c.t
\end{split}
\end{equation}
We then go through the successive changes of variables
$(u,s) \to (u, y=|e|Bsu)$, $t=iy$, last $z=ut$, plus a Wick rotation (see
subsection \ref{subsec:dm0} below),
 to get
\begin{equation}
\hskip -5mm
\Sigma^{n=0}(\pi_0,\pi_\perp)  \stackrel{B\to 0}{\sim}
\frac{\alpha m}{\pi} \int_0^\infty dz
\int_0^1\frac{du}{u^2\sqrt{1-u}}\;
 e^{\displaystyle -\frac{zm^2}{u|e|B}}\;e^{\displaystyle
z\frac{1-u}{u}\frac{\pi_0^2}{|e|B}\;\pi_0^2}\;
e^{\displaystyle -z \frac{1-u}{u^2}\;\frac{\pi_\perp^2}{|e|B}}
+\ terms\ \propto\ (1+i\gamma_1\gamma_2) + c.t
\end{equation}
If we now go on mass-shell, $\pi\!\!\!/ +m=0$, $\pi\!\!\!/^2 = m^2= -\pi^2
-\frac{|e|}{2} 2 \sigma^3 B \Rightarrow m^2 = \pi_0^2 -\pi_\perp^2
-|e|\sigma^3 B$, we get
\begin{equation}
\hskip -1cm
\Sigma^{n=0}_{mass-shell}(\pi_0,\pi_\perp)  \stackrel{B\to 0}{\sim}
\frac{\alpha m}{\pi} \int_0^\infty dz
\int_0^1\frac{du}{u^2\sqrt{1-u}}\;
e^{\displaystyle -z\frac{m^2}{|e|B}}\; e^{\displaystyle z\frac{1-u}{u}\sigma^3}\;
e^{\displaystyle -z\frac{(1-u)^2}{u^2}\frac{\pi_\perp^2}{|e|B}}
+\ terms\ \propto\ (1+i\gamma_1\gamma_2) + c.t
\end{equation}
The 1st renormalization condition in (\ref{eq:rencond})
 concerns the vanishing, on mass-shell, of
$\Sigma$ at the limit $B\to 0$. We have therefore to introduce a 1st
counterterm $c.t._1$
\begin{equation}
c.t._1 = -\lim_{B\to 0}\frac{\alpha m}{\pi} \int_0^\infty dz
\int_0^1\frac{du}{u^2\sqrt{1-u}}\;
e^{\displaystyle -z\frac{m^2}{|e|B}}\; e^{\displaystyle
z\frac{1-u}{u}\sigma^3}\;
e^{\displaystyle -z\frac{(1-u)^2}{u^2}\frac{\pi_\perp^2}{|e|B}}
+\ terms\ \propto\ (1+i\gamma_1\gamma_2)
\label{eq:LLct1}
\end{equation}
(the terms $\propto(1+i\gamma^1 \gamma^2)$ give vanishing contribution
only to external LLL).

The second renormalization condition (see (\ref{eq:rencond}))
 concerns the derivative of $\Sigma$.
This leads to introducing a second set of counterterms. However, they have
to vanish on mass-shell since they must satisfy the 1st renormalization
condition. Since, in order to calculate $\delta m$, we precisely work on
mass-shell, we can forget about the second set of counterterms and proceed
now with the calculation of $\delta m^0_{LLL}$.

\subsection{Calculation of the 1-loop self-mass
$\boldsymbol{\delta m^0_{LLL}}$
when both external and internal electrons are in the lowest Landau level}
\label{subsec:dm0}

When acting on external LLL electrons, and on mass-shell, one has
$\pi_0^2=m^2, \pi_\perp^2= \sigma^3 eB= -eB = +|e|B$. 
From (\ref{eq:LLspace}) and (\ref{eq:LLct1}) one then obtains
\begin{equation}
\hskip -15mm
\delta m^0_{LLL} = \frac{\alpha m}{\pi} \int_0^\infty ds \int_0^1
\frac{du}{\sqrt{1-u}}\;\Big[\cos \tan^{-1}
\frac{|e|Bs(1-u)}{1+i|e|Bs(1-u)}\Big]\frac{i|e|B}{1+i|e|Bs(1-u)}
 e^{\displaystyle -is u^2 m^2} e^{\displaystyle -i\tan^{-1}
\frac{|e|Bs(1-u)}{1+i|e|Bs(1-u)}}+c.t._1
\label{eq:del0}
\end{equation}

After some calculations which are detailed in appendix
\ref{section:delmLLL}, one gets

\begin{equation}
\hskip -15mm
\delta m^0_{LLL}=
\frac{\alpha m}{4\pi}\;
\underbrace{4\int_0^{\infty} dz \int_0^1 \frac{du}{\sqrt{1-u}}
\Big[\cosh \tanh^{-1}\, \frac{z(1-u)}{u^2+z(1-u)}\Big]
\frac{1}{u^2+z(1-u)}\;e^{-zm^2/|e|B}\;
e^{\displaystyle -\tanh^{-1}\, \frac{z(1-u)}{u^2+z(1-u)}}}_{\hat I^0(L),
L=|e|B/m^2}+c.t._1
\label{eq:delm0LLL}
\end{equation}

When $m=0$, $\delta m^0_{LLL}= \frac{\alpha m}{\pi}\int_0^{\infty} dz
\int_0^1 \frac{du}{\sqrt{1-u}}\;\frac{1}{u^2+z(1-u)}\;\frac12\;\Big( 1+
e^{-2 \tanh^{-1}
\frac{z(1-u)}{u^2+z(1-u)}}\Big) \sim \frac{\alpha m}{2\pi}\int_0^{\infty}
dz \int_0^1 \frac{du}{\sqrt{1-u}}\;\frac{1}{u^2+z(1-u)}$ which diverges at
$z \to \infty$. Like before, one must eventually take the limit $m \to 0$
after the integration has been performed.

The exponential $e^{-2 \tanh^{-1} \frac{z(1-u)}{u^2+z(1-u)}}$  being bounded
by $1$ and going to $0$
when $z \to \infty$, we have  to evaluate
\begin{equation}
\delta m^0_{LLL} \sim
\frac{\alpha m}{4\pi}\; \underbrace{2\int_0^{\infty} dz
\int_0^1 \frac{du}{\sqrt{1-u}}\;\frac{e^{-zm^2/|e|B}}{u^2+z(1-u)}\ (1 +
\epsilon)}_{\hat I^0(L), L=|e|B/m^2}+c.t._1, \quad 0\leq \epsilon\leq 1,
\end{equation}
in which we have, like previously, factorized $\frac{\alpha m}{4\pi}$, at
the price
of introducing an extra factor $2$ in front of the integral.
One accordingly defines now (compare with (\ref{eq:I2}) (\ref{eq:Iapp2})) 
\begin{equation}
\begin{split}
\hat I^0(L) &=4\int_0^{\infty} dz \int_0^1 \frac{du}{\sqrt{1-u}}
\Big[\cosh \tanh^{-1}\, \frac{z(1-u)}{u^2+z(1-u)}\Big]
\frac{1}{u^2+z(1-u)}\;e^{-zm^2/|e|B}\;
e^{\displaystyle -\tanh^{-1}\, \frac{z(1-u)}{u^2+z(1-u)}}\cr
& \simeq 2\int_0^{\infty} dz
\int_0^1 \frac{du}{\sqrt{1-u}}\;\frac{e^{-zm^2/|e|B}}{u^2+z(1-u)}\ (1 +
\epsilon),\quad 0\leq \epsilon \leq 1.
\end{split}
\end{equation}
Note that, unlike when taking all Landau levels of the internal electrons
 into account, the integral $\hat I_0(L)$ is convergent at $z=0$ without
introducing any counterterm.

One has
\begin{equation}
\hskip -1cm
g_0(z)\equiv \int_0^1 \frac{du}{\sqrt{1-u}}\;\frac{2}{u^2+z(1-u)}=
-\frac{4 \sqrt{2} \left(\displaystyle\frac{\tan
^{-1}\left(\displaystyle\frac{\sqrt{2}}{\sqrt{z+\sqrt{(z-4)
z}-2}}\right)}{\sqrt{z+\sqrt{(z-4) z}-2}}-\displaystyle\frac{\tan
^{-1}\left(\displaystyle\frac{\sqrt{2}}{\sqrt{z-\sqrt{(z-4)
z}-2}}\right)}{\sqrt{z-\sqrt{(z-4) z}-2}}\right)}{\sqrt{(z-4) z}},
\label{eq:g0}
\end{equation}
(to be compared with (\ref{eq:g}))
such that
\begin{equation}
\delta m^0_{LLL} \sim \frac{\alpha m}{4\pi} \int_0^\infty
dz\;e^{-zm^2/|e|B}\;g_0(z) + c.t._1.
\label{eq:delmg0}
\end{equation}

\subsubsection{Contribution of the counterterm to $\boldsymbol{\delta
m^0_{LLL}}$}

For external LLL, $\pi_\perp^2 \to |e|B, \sigma^3 \to -1$, this counterterm
contributes to $\delta m^0_{LLL}$ by
\begin{equation}
c.t._1=-\lim_{B\to 0}
\frac{\alpha m}{\pi} \int_0^\infty dz
\int_0^1\frac{du}{u^2\sqrt{1-u}}\;e^{\displaystyle -z\frac{m^2}{|e|B}}\;
e^{\displaystyle -z\frac{1-u}{u^2}},
\end{equation}
which is convergent.  It yields
\begin{equation}
c.t._1=-\lim_{B\to 0}
\frac{\alpha m}{\pi}\int_0^1\frac{du}{u^2\sqrt{1-u}}\;
\frac{1}{\frac{m^2}{|e|B} + \frac{1-u}{u^2}}
=- \lim_{B\to
0}\frac{\alpha}{2\pi}\;\frac{|e|B}{m}\;g_0\left(\frac{|e|B}{m^2}\right),
\label{eq:ct1}
\end{equation}
in which $g_0$ is the same as that defined in (\ref{eq:g0}).
At the limit $z\to 0$
\begin{equation}
g_0(z) \stackrel{z\to 0}{\sim}
\frac{\pi}{\sqrt{z}} + 2\ln 2-\frac{\ln z}{2}
+\frac{z}{16} (-\ln z-1+4 \ln 2) +{\cal O}(z^{3/2}),
\label{eq:limg0}
\end{equation}
such that
\begin{equation}
c.t._1 = -\lim_{B\to 0} \Big(\frac{\alpha}{2}\;\sqrt{|e|B} +\frac{\alpha}{\pi}\;
\frac{|e|B}{m}\;\ln 2\Big) + \ldots
\end{equation}
which we shall truncate at the first term since the limit $m\to 0$ should
be taken afterwards. Accordingly, one finds a vanishing counterterm (which
is in particular independent of the external $B$)
\begin{equation}
c.t._1 =0.
\label{eq:ct1final}
\end{equation}

Collecting (\ref{eq:delmg0}) and (\ref{eq:ct1final}) yields
\begin{equation}
\delta m^0_{LLL} \sim \frac{\alpha m}{4\pi}
\int_0^\infty dz\;e^{-zm^2/|e|B}\;g_0(z).
\label{eq:deltam0}
\end{equation}

Notice that the bare $\delta m^0_{LLL}$ (and, of course, the (vanishing)
counterterm) are both finite, unlike  when all Landau levels of the internal
electron are accounted for.

\subsection{The limit of $\boldsymbol{\delta m^0_{LLL}}$ when
$\boldsymbol{m\to 0}$}

In addition to the limit $z\to 0$ given in (\ref{eq:limg0}) one has
\begin{equation}
g_0(z) \stackrel{z\to \infty}{\sim} \frac{2\pi}{\sqrt{z}}
-\frac{4}{z} + \frac{2\pi}{z^{3/2}}-\frac{32}{3z^3} +\ldots.
\label{eq:gliminf}
\end{equation}
So, splitting the $z$ interval of integration of (\ref{eq:deltam0})
 into 3  sub-intervals gives
\begin{equation}
\hskip -1cm
\delta m^0_{LLL} \sim \frac{\alpha m}{4\pi}
\Bigg[
\int_0^a dz\; e^{-z m^2/|e|B} \frac{\pi}{\sqrt{z}}
+\underbrace{\int_a^b dz\; e^{-z m^2/|e|B} g_0(z)}_{constant}
+\int_b^\infty dz\; e^{-z m^2/|e|B} \frac{2\pi}{\sqrt{z}}\quad
\Bigg].
\end{equation}
The bounds $a$ and $b$ are chosen such that, for $z\in[0,a]$ the expansion
(\ref{eq:limg0}) is valid, and for $z\in [b,\infty]$ the expansion
(\ref{eq:gliminf}) is valid.
Since
\begin{equation}
\begin{split}
\int dz\; \frac{e^{-z m^2/|e|B}}{\sqrt{z}} &= \frac{\sqrt{\pi}
\; Erf\left(\sqrt{m^2/|e|B} \sqrt{z}\right)}{\sqrt{m^2/|e|B}},
\end{split}
\end{equation}
one has
\begin{equation}
\begin{split}
\delta m^0_{LLL} & \sim \frac{\alpha m}{4\pi}
\Bigg[
\pi\frac{\sqrt{\pi}}{\sqrt{m^2/|e|B}}
\Big(Erf(\sqrt{m^2/|e|B}\sqrt{a}) -
\frac{4}{\sqrt{\pi}}\times 0\Big)\cr
&   + cst \cr
& + 2\pi\;\frac{\sqrt{\pi}}{\sqrt{m^2/|e|B}}
\Big(\underbrace{Erf(\sqrt{m^2/|e|B}\sqrt{z=\infty})}_{1}-
Erf(\sqrt{m^2/|e|B}\sqrt{b})\Big)\quad
\Bigg].
\end{split}
\end{equation}
To study the limit $m\to 0$ we use 
\begin{equation}
\begin{split}
Erf(x) & \stackrel{x\to 0}{\sim} \frac{2x}{\sqrt{\pi}},\cr
Erf(x) & \stackrel{x\to\infty}{\sim} 1,
\end{split}
\end{equation}
which shows that it is the value at $z=\infty$ that controls $\delta
m^0_{LLL}$.

Finally
\begin{equation}
\delta m^0_{LLL} \stackrel{m\to 0}{\to} \frac{\alpha}{2}\; \sqrt{\pi\,|e|B}
=\sqrt{2}\;\delta m_{LLL}.
\label{eq:delm0lim}
\end{equation}

\subsection{An approximate analytical expression for $\boldsymbol{\delta
m^0_{LLL}}$. Comparison with $\boldsymbol{\delta m_{LLL}}$}

It is easy to get a fair approximate analytical expression for $\delta
m^0_{LLL}$ given in (\ref{eq:deltam0})
by using the following simple fit to $g_0(z)$ 
\begin{equation}
g_0^{app}(z) \simeq e^{-z/30}\Big( \frac{\pi}{\sqrt{z}} + 2\ln 2 \Big)
+e^{-30/z}\Big(\frac{2\pi}{\sqrt{z}}-\frac{4}{z}\Big),
\label{eq:g0app}
\end{equation}
which has, in particular, the appropriate limits at $z \to 0$ and $z \to
\infty$.
On fig.~5 the exact $g_0$ is plotted in blue and the approximate one in
yellow.

\vbox{
\begin{center}
\includegraphics[width=6cm, height=4cm]{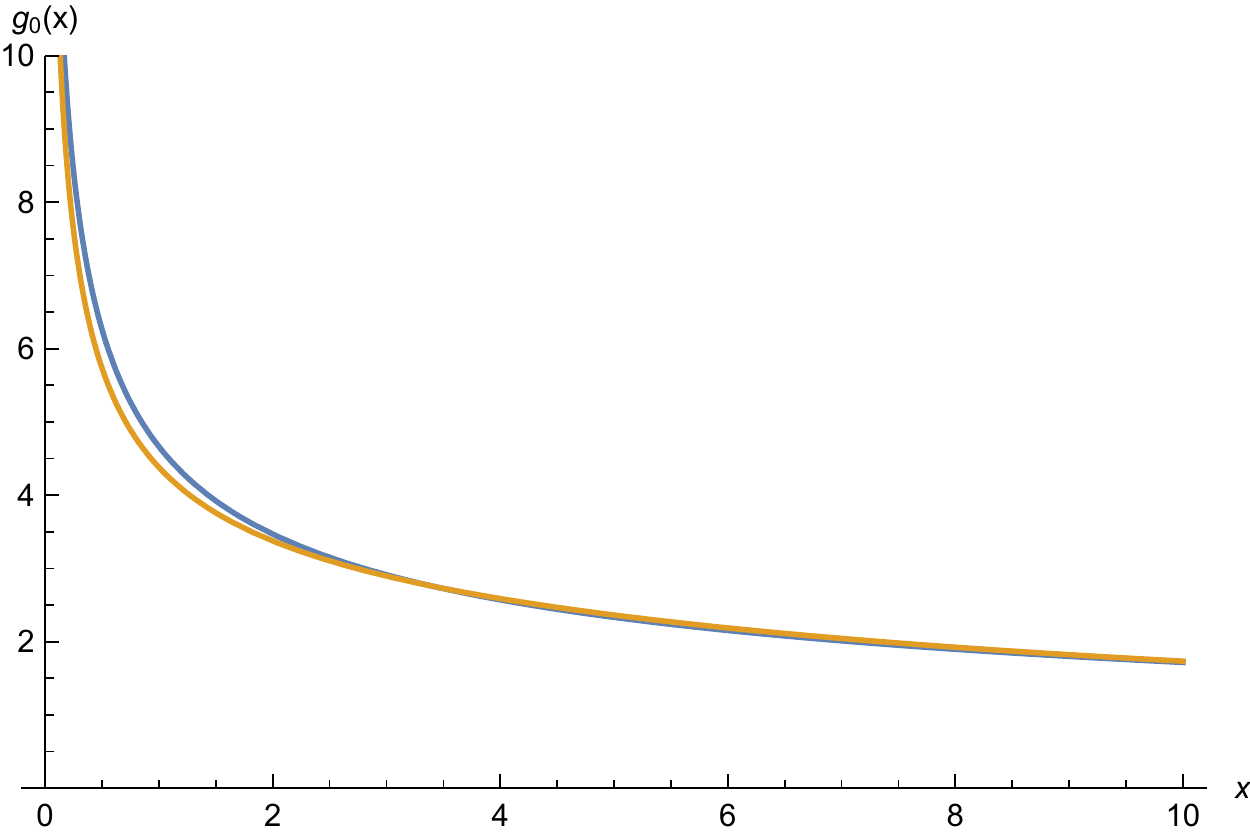}

\smallskip
{\em Fig.~5: the exact $g_0$ given in (\ref{eq:g0}  (blue) and its
approximate expression (\ref{eq:g0app}) (yellow) }
\end{center}
}

This yields
\begin{equation}
\delta m^0_{LLL} \approx \frac{\alpha m}{4\pi}\;\left(\frac{2\pi ^{3/2}\;
\exp\Big[\displaystyle-2 \sqrt{30} \sqrt{\frac{m^2}{|e|B}}\Big]}{\sqrt{\displaystyle\frac{m^2}{|e|B}}}
+\frac{\pi ^{3/2}}{\sqrt{\displaystyle\frac{m^2}{|e|B}+\displaystyle\frac{1}{30}}}
+\frac{60 \ln (2)}{30\, \displaystyle\frac{m^2}{|e|B}+1}
-8\, BesselK\big(0,2 \sqrt{30} \sqrt{\frac{m^2}{|e|B}}\big)\right),
\label{eq:dm0ana}
\end{equation}
which has the limit (\ref{eq:delm0lim}) when $m\to 0$. Notice also that the
second contribution yields a finite $\delta m^0_{LLL}\to
\frac{\alpha}{4\pi}\; \pi^{3/2}\sqrt{|e|B}$ when $m \to \infty$. 

On Fig.~6, we plot $\sqrt{\frac{m^2}{|e|B}}\int_0^\infty
dz\;e^{-zm^2/|e|B}\;g_0(z)$ in blue together with
$\sqrt{\frac{m^2}{|e|B}}\int_0^\infty
dz\;e^{-zm^2/|e|B}\;g_0^{app}(z)$ in yellow, which corresponds to $4\pi
\delta m^0_{LLL}/\alpha \sqrt{|e|B}$. It shows that this rather
crude approximation is good at better than $7\%$ for $\frac{m^2}{|e|B} \geq .4$, at $\sim
10\%$ for lower values of $\frac{m^2}{|e|B}$ and that it has, of course, the
appropriate limit $2\pi^{3/2} \approx 11.14$ at $\frac{m^2}{|e|B} =0$.

\vbox{
\begin{center}
\includegraphics[width=6cm, height=4cm]{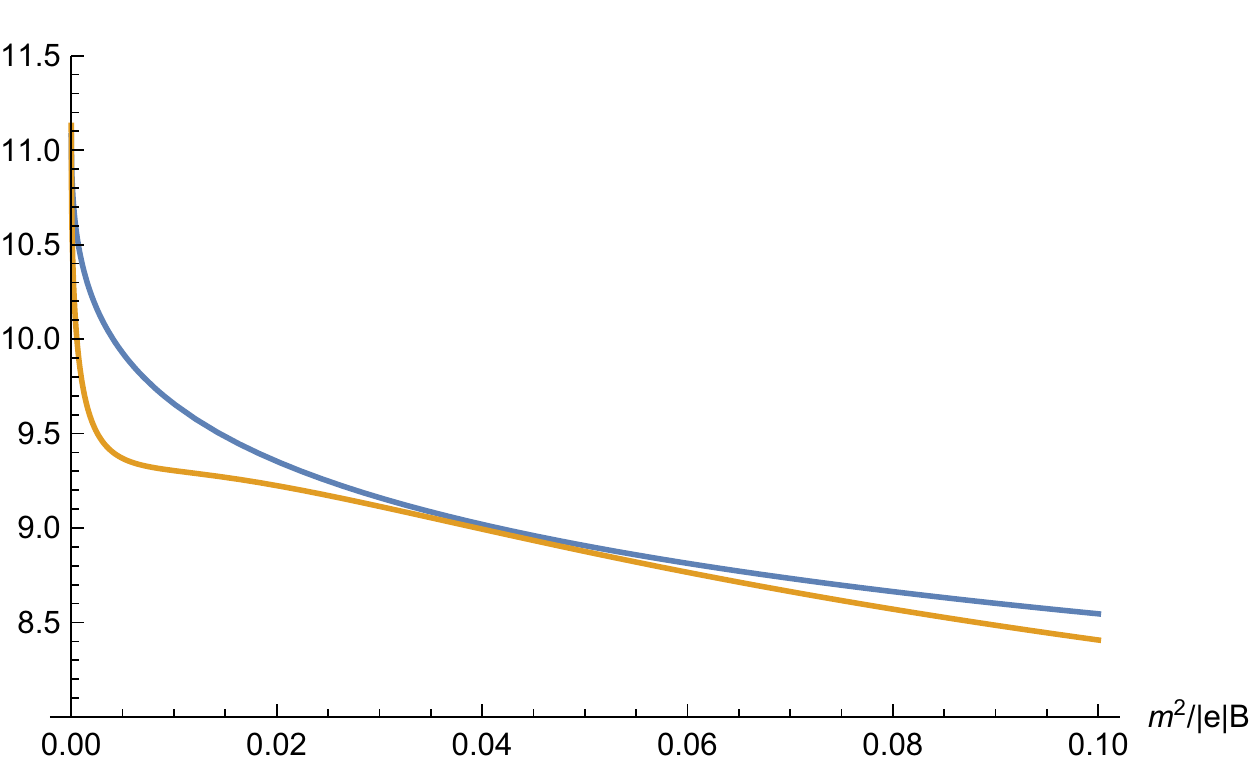}
\hskip 2cm
\includegraphics[width=6cm, height=4cm]{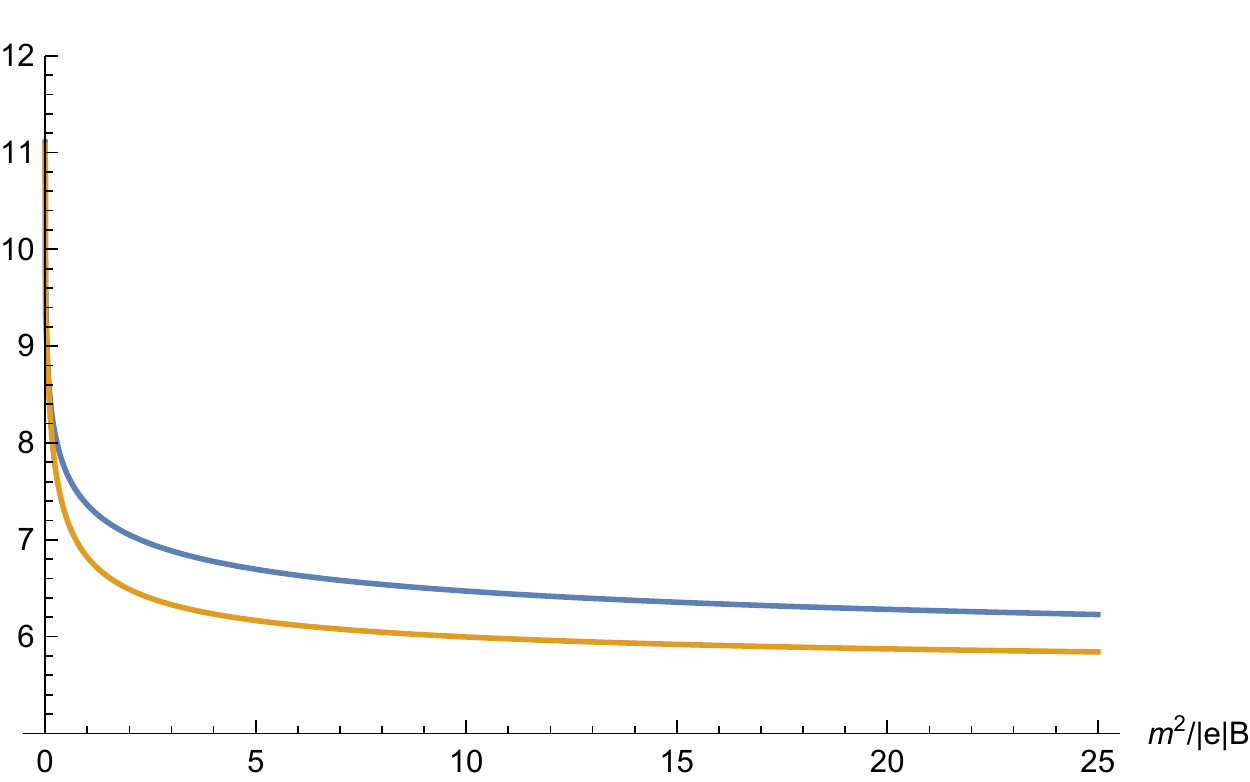}

\smallskip
{\em Fig.~6: $\sqrt{\frac{m^2}{|e|B}}\int_0^\infty
dz\;e^{-zm^2/|e|B}\;g_0(z)$ in blue and
$\sqrt{\frac{m^2}{|e|B}}\int_0^\infty
dz\;e^{-zm^2/|e|B}\;g_0^{app}(z)$ in yellow as functions of
$\frac{m^2}{|e|B}$. }
\end{center}
}

Fig.~6 also shows that this approximation is the worse in the close vicinity of
$\frac{m^2}{|e|B} =0$.
Including higher orders in the expansions of $g_0(z)$ at $z\to 0$ and $z\to
\infty$ turns out to improve the situation at large values of
$\frac{m^2}{|e|B}$ but,
instead, to worsen it close to $0$.

On Fig.~7 we plot $\frac{4\pi}{\alpha}\frac{ \delta m_{LLL}}{\sqrt{|e|B}}$ given in
(\ref{eq:delmg}) and (\ref{eq:g}) in blue together with
$\frac{4\pi}{\alpha}\frac{ \delta m^0_{LLL}}{\sqrt{|e|B}}$ given in
(\ref{eq:deltam0}) and (\ref{eq:g0})
 as functions of $\frac{m^2}{|e|B}$. They determine the behavior of the
corresponding $\delta m$'s  at fixed value of $|e|B$ when $m$ becomes
larger and larger (and not their limits at $|e|B \to 0$, which vanishes for
both in virtue of the first renormalization condition).
As we see, this behavior is very different for the two cases:
$\frac{4\pi}{\alpha}\frac{\delta m_{LLL}}{\sqrt{|e|B}}$ behaves
like $\frac{e^{-m^2/|e|B}}{\sqrt{m^2/|e|B}} \to
0$ when $\frac{m^2}{|e|B} \to \infty$ while $\frac{4\pi}{\alpha}\frac{\delta
m^0_{LLL}}{\sqrt{|e|B}}$ goes  to $\pi^{3/2}$ at the same limit.

\vbox{
\begin{center}
\includegraphics[width=6cm, height=4cm]{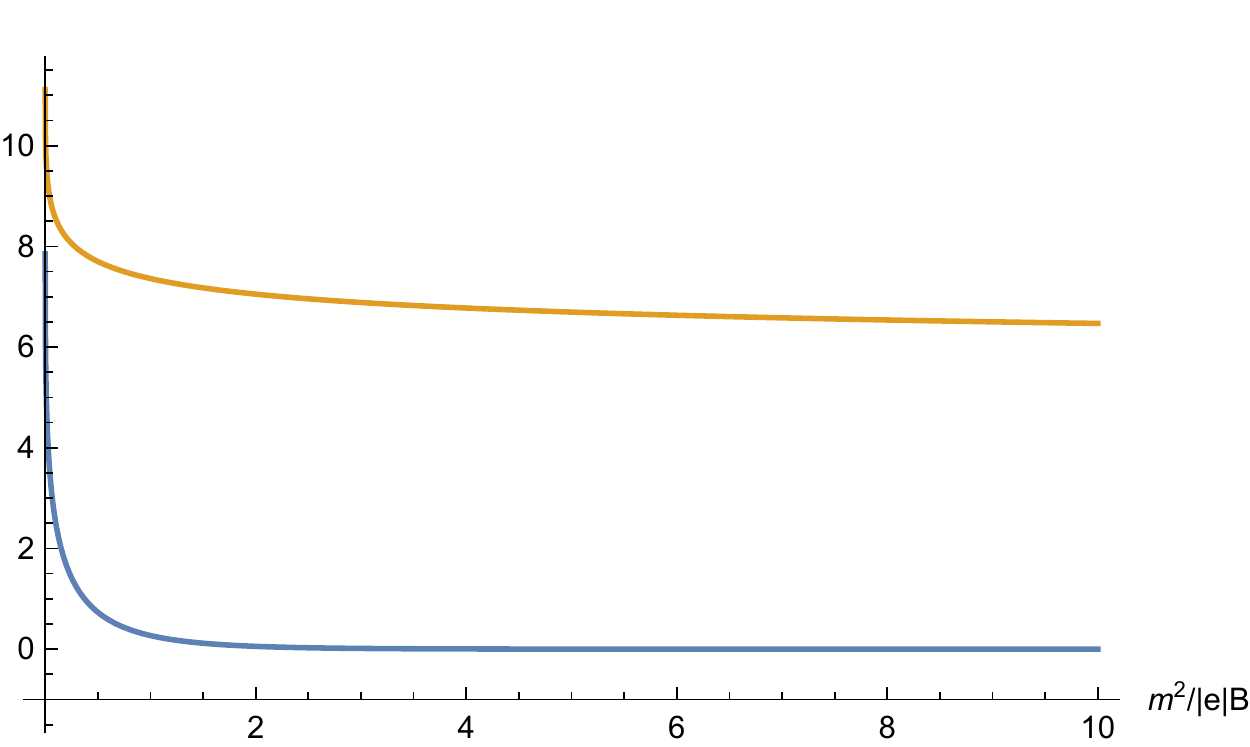}

\smallskip
{\em Fig.~7: $\frac{4\pi}{\alpha}\frac{\delta m_{LLL}}{\sqrt{|e|B}}$ given in
(\ref{eq:dm3}) (blue) and
$\frac{4\pi}{\alpha}\frac{\delta m^0_{LLL}}{\sqrt{|e|B}}$ given in (\ref{eq:dm0ana})
 (yellow) as functions of $\frac{m^2}{|e|B}$ }
\end{center}
}

On Fig.~8
\footnote{Figs.~7 and 8 are not plotted with the approximate analytical
expressions that we have deduced for the $\delta m$'s, but by numerical
integration of their exact expressions.}
 we now plot $\frac{4\pi}{\alpha}\frac{\delta m_{LLL}}{m}$ (in blue) and
$\frac{4\pi}{\alpha }\frac{\delta m^0_{LLL}}{m}$ (in yellow) as functions of
$\frac{|e|B}{m^2}$. This shows
how the $\delta m$'s vary with $B$ at fixed $m$. Once more, while we
witness as expected their both vanishing at $B=0$ according to the 1st
renormalization condition, their behavior $\propto \sqrt{|e|B}$  when $B$
becomes larger and larger is factorized by different coefficients; as a
result  $\delta m^0_{LLL}/m$ is already more than twice $\delta m_{LLL}/m$
at $\frac{|e|B}{m^2}=20$.
Restricting the internal
electron to its LLL results accordingly in a very large overestimate of the
self-mass.
 
\vbox{
\begin{center}
\includegraphics[width=6cm, height=4cm]{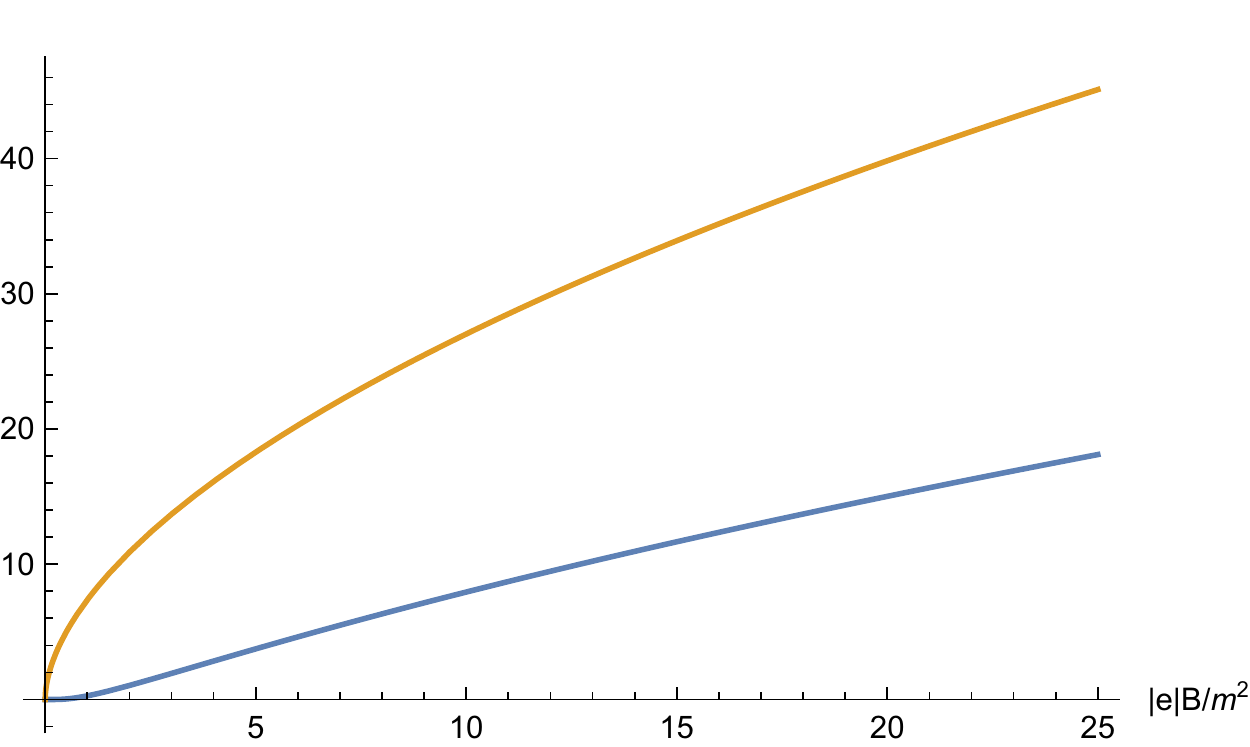}

\smallskip
{\em Fig.~8: $\frac{4\pi}{\alpha}\frac{\delta m_{LLL}}{m}$ (blue) and
$\frac{4\pi}{\alpha}\frac{\delta m^0_{LLL}}{m}$ (yellow) as functions of
$\frac{|e|B}{m^2}$}
\end{center}
}

\subsection{A few remarks}

$\delta m_{LLL}$ and $\delta m^0_{LLL}$ do not have the
same limits at $m\to 0$, nor at $m\to\infty$.

 Would $m\to 0$ be equivalent to $eB \to \infty$,
one could, at first sight, expect that only the LLL plays a role.
This would however only be true
if the only physical variable was $|e|B/m^2$, and if renormalization did
not put a grain of salt in such an argumentation.

While it is  true that
$G^{n=0}(p,B)$  can indeed be obtained by formally taking the limit
$B\to\infty$ of $G_0(p,B)$ (see Appendix \ref{section:LLLprop}), one should notice that:

* this limit cannot be applied to the phase $\Phi$;

* the factor $e^{-k_\perp^2/|e|B}$ is not replaced by $1$ inside $G^{n=0}$
despite $B\to \infty$; this is because, as the Larmor radius shrinks to $0$
at this limit, $k_\perp$ can extend to $\infty$;

* the (vanishing) counterterm is determined by taking first the limit $B\to 0$, so as
to fulfill
renormalization conditions; then, eventually, the non-vanishing limit $m\to
0$ is taken; therefore,  naively taking the limit $B\to \infty$ to ``select'' the
LLL cannot be applied either to the counterterm.

Arguing that the limit $m\to 0$ is
equivalent to $B\to \infty$ can accordingly only be wrong 
\footnote{Eventually forcing the identity between the two limits at $m=0$
of $\delta m_{LLL}$ and $\delta m^0_{LLL}$ as a kind of renormalization
condition must be rejected.}.

The limits at $m\to \infty$ (which should not be confused with those at
$B\to 0$) are also very different since $\delta m_{LLL}
\sim \frac{|e|B}{m}\;e^{-m^2/|e|B}\to 0$ while $\delta m^0_{LLL} \sim cst
\times \sqrt{|e|B}$ (see Fig.~7).

Large cancellations therefore occur among multiple Landau levels of the
virtual electron. However,
they  can only be estimated after going  through the filter of
renormalization, and infinities that need being tamed
 only arise when one accounts for all levels.

\section{Conclusion and prospects}

Unlike what happens for QED$_{3+1}$, the massless limit of the 1-loop
$\delta m_{LLL}$ in external $B$ for QED$_{3+1}$ reduced on a 2-brane
does not vanish. We have shown furthermore that it corresponds
to  an electron propagating inside a graphene-like medium.
The latter cannot therefore stay ``gapless'' at 1-loop in the
presence of a magnetic field.
This result has been obtained with special attention paid to the
renormalization conditions.

The result is very simple because we have restricted the
external electron to lie in the lowest Landau level. For higher
levels, the situation is much more intricate and analytical formul{\ae}
certainly cannot be obtained.

We have also shown that restricting to the LLL of the internal electron
largely overestimates the self-mass; in particular, its value when $m\to 0$
triggers a multiplicative factor $\sqrt{2}$. Despite the case under concern has
the peculiarity that taming infinities and renormalizing is only needed
when accounting for all Landau levels,  studies based on such an
approximation  appear rather suspicious.
Note that, in the case of standard
QED$_{3+1}$, it was shown in \cite{Machet2} that accounting for the sole
leading $(\ln)^2$ terms largely increases the result, too. 

I cannot pretend to have dealt with real graphene, in which, in particular,
the smallness of the Fermi velocity with respect to the speed of light
makes the theory strongly coupled. There, techniques have to be mastered
which go beyond perturbative expansions, while respecting appropriate
renormalization conditions.

It is also well known that the photon propagator gets modified
in the presence of an  external $B$ (see for example \cite{MachetVysotsky2011}).
This modification has been included in calculations of the electron
self-energy \cite{LS1983} \cite{KMO-MPLA-2002} with the result that double
logs are turned into single logs. However, the large single logs closely
associated with counterterms (see \cite{Machet1})
 were not taken into account. Furthermore, this
modification of the photon propagator and the eventual screening of the
Coulomb potential is obtained by resumming the infinite geometric series
of 1-loop vacuum polarizations (see for example \cite{MachetVysotsky2011});
in contrast, Quantum Field Theory stipulates that renormalization conditions
and the addition of the corresponding counterterms should be achieved
consistently order by order in powers of the coupling constant or in the
number of loops. In this framework, screening the Coulomb potential inside
the electron self-energy at finite order raises many issues, both technical
and conceptual.

\vskip .5cm

{\em \underline{Acknowledgments}: It is a great pleasure to thank
M.I.~Vysotsky for his invaluable assistance and advice.}

\newpage

\appendix

{\bf\Large Appendix}

\section{The  propagator $\boldsymbol{G^{n=0}(p,B)}$  of an
electron in the lowest Landau level as the limit at $\boldsymbol{B\to
\infty}$ of $\boldsymbol{G_0(p,B)}$ (without the phase (\ref{eq:phase}))}
\label{section:LLLprop}

After putting aside the phase $\Phi$ given in (\ref{eq:phase}),
we can get it by taking the limit $B\to \infty$ in $G(p,B)$

Let us  consider the general expression (6) of
\cite{Tsai1974} ($z=|e|Bs_1$), which does not include the phase
\begin{equation}
-iG_0(p,B)= \int_0^\infty ds_1 e^{\displaystyle -is_1[m^2-i\epsilon
+p_\parallel^2 +\frac{\tan z}{z}p_\perp^2]}
\frac{e^{iq\sigma^3 z}}{\cos z}\Big(m-(\gamma p)_\parallel
-\frac{e^{-iq\sigma^3 z}}{\cos z}(\gamma p)_\perp\Big).
\end{equation}
Since $(q\sigma_3)^2=1$, $\cos q\sigma^3 z = \cos z$ and $\sin q\sigma^3 z =
q\sigma^3 \sin z$.
As $\sigma^3 = i \gamma^1 \gamma^2$,
if one cancels at the beginning the 2 inverse exponentials one gets
\begin{equation}
-iG_0(p,B)= \int_0^\infty ds_1\; e^{\displaystyle -is_1\big(m^2-i\epsilon
+p_\parallel^2 +\frac{\tan z}{z}p_\perp^2\big)}
\Big(\big(1 -q\gamma^1 \gamma^2 \frac{\sin z}{\cos z}\big)
\big(m-(\gamma p)_\parallel\big)-\frac{(\gamma p)_\perp}{\cos^2 z}\Big).
\end{equation}
To take the limit $B\to \infty$ one must first make a Wick rotation
$s_1=-iy_1$. Then, $\sin z=-i\sinh |e|By_1, \cos z = \cosh |e|By_1$ and
\begin{equation}
\begin{split}
& \hskip -1.5cm
-iG_0(p,B)= -i\int_0^{i\infty} dy_1\; e^{\displaystyle -y_1\big(m^2-i\epsilon
+p_\parallel^2 +\frac{-i\tanh |e|By_1}{-i|e|By_1}p_\perp^2\big)}
\Big(\big(1 -q\gamma^1 \gamma^2 \frac{-i\sinh |e|By_1}{\cosh |e|By_1}\big)
\big(m-(\gamma p)_\parallel\big)-\frac{(\gamma p)_\perp}{\cosh^2
eBy_1}\Big)\cr
&=-i\int_0^{i\infty} dy_1\; e^{\displaystyle -y_1\big(m^2-i\epsilon
+p_\parallel^2 +\frac{\tanh |e|By_1}{|e|By_1}p_\perp^2\big)}
\Big(\big(1 +iq\gamma^1 \gamma^2 \frac{\sinh eBy_1}{\cosh eBy_1}\big)
\big(m-(\gamma p)_\parallel\big)-\frac{(\gamma p)_\perp}{\cosh^2
eBy_1}\Big).
\end{split}
\end{equation}
Then, $\int_0^{i\infty} + \int_{1/4\; circle}+\int_\infty^0 dy_1 = \sum
residues$. If we suppose that $\int_{1/4\; circle} =0$ and that $\sum
residues=0$, $\int_0^{i\infty} dy_1 = \int_0^\infty dy_1$ and
\begin{equation}
-iG_0(p,B)=-i\int_0^{\infty} dy_1\; e^{\displaystyle -y_1\big(m^2-i\epsilon
+p_\parallel^2 +\frac{\tanh |e|By_1}{|e|By_1}p_\perp^2\big)}
\Big(\big(1 +iq\gamma^1 \gamma^2 \frac{\sinh eBy_1}{\cosh eBy_1}\big)
\big(m-(\gamma p)_\parallel\big)-\frac{(\gamma p)_\perp}{\cosh^2
eBy_1}\Big),
\end{equation}
on which we can now take the limit $B\to \infty$.
\begin{equation}
-iG_0(p,B) \stackrel{B\to \infty}{\to}
-ie^{-p_\perp^2/|e|B}\int_0^{\infty} dy_1\;
e^{\displaystyle -y_1(m^2 +p_\parallel^2)}\;
\Big((1 +iq\gamma^1 \gamma^2)
\big(m-(\gamma p)_\parallel\big)
\Big).
\end{equation}
This is the usual result (\ref{eq:GLLL}) for $G^{n=0}(p,B)$ since $q=-1$.

If we had used instead eq.~(2.47b) of  \cite{DittrichReuter}, in
which $e<0$, we would have got the wrong projector $1+i\gamma^1\gamma^2$,
while, with their conventions, the wave function of the LLL is the same.
The exponentials $e^{\pm iz\sigma^3}$ of
\cite{DittrichReuter},
which should in reality be $e^{\pm iq z \sigma^3}$ with $q=-1$.
This is one of the rare examples in QED where the sign of the electric charge matters.

\section{Demonstration of (\ref{eq:I2})}
\label{section:DemeurJanco}

In (\ref{eq:I0}) it is interesting to expand  $e^{i\beta}$ into $\cos\beta +
i\sin\beta$ and to use the expressions (\ref{eq:beta})
 of $\cos\beta$ and $\sin\beta$  to cast $\delta m$ in the form
\begin{equation}
\begin{split}
\hat I &=\int_0^\infty \frac{ds}{s}\int_0^1
\frac{du}{\sqrt{1-u}}\;e^{-iuy\frac{m^2}{|e|B}}
\Big[\frac{e^{-iqy}\big((1-u)\cos y +u \sin y/y +i(1-u)\sin
y\big)}{\Delta(u,y)}\big(1+ue^{2iqy}\big)-(1-u)\Big]\cr
 &= \int_0^\infty \frac{ds}{s}\int_0^1
\frac{du}{\sqrt{1-u}}\;e^{-iuy\frac{m^2}{|e|B}}
\Big[\frac{1-u+u \sin y/y\;e^{-iqy}}{\Delta(u,y)} \big(1+ue^{2iqy}\big)-(1-u)\Big]
\end{split}
\end{equation}
then to notice that $\Delta(u,y) = (1-u +u\frac{\sin y}{y}\,e^{+iqy})(1-u
+u\frac{\sin y}{y}\,e^{-iqy})$ to simplify the previous expression into
\begin{equation}
\hat I =\int_0^\infty \frac{ds}{s}\int_0^1
\frac{du}{\sqrt{1-u}}\;e^{-iuy\frac{m^2}{|e|B}}\;
\Big[\frac{1+ue^{2iqy}}{1-u+u\frac{\sin y}{y}\,e^{+iqy}}
-(1+u)\Big]
\end{equation}
After the change of variables
(we shall come back later to this change of variables which introduces in
particular a dependence of the counterterm on $L$)
\begin{equation}
(u,s)\to (u,y=|e|Bsu) \Rightarrow \frac{du\, ds}{s}=\frac{du\, dy}{y}
=du\,\frac{d(qy)}{qy},
\label{eq:chvar1}
\end{equation}
it becomes
\begin{equation}
\hat I =\int_0^{q\infty} \frac{d(qy)}{qy}\int_0^1
\frac{du}{\sqrt{1-u}}\;e^{-iuy\frac{m^2}{|e|B}}\;
\Big[\frac{1+ue^{2iqy}}{1-u+u\frac{\sin qy}{qy}\,e^{+iqy}}
-(1+u)\Big]
\end{equation}
Noticing that, since $q=-1$, $\sin y/y =\sin qy/qy$ and expressing  $\sin qy$ in
the denominator in terms of complex exponentials gives
\begin{equation}
\hat I(L) = \int_0^{q\infty} d(qy)
\int_0^1 \frac{du}{\sqrt{1-u}}\; e^{-iuy\frac{m^2}{|e|B}}
\left[ \frac{2i\left(1+u\,e^{2iqy}\right)}{2iqy(1-u)
+u\left(e^{2iqy}-1\right)}
-\frac{1+u}{qy}
\right].
\end{equation}
Going to $t=-iqy$ yields
\begin{equation}
\hat I(L)= \int_0^{-iq\infty} dt
\int_0^1 \frac{du}{\sqrt{1-u}}\; e^{uqt\frac{m^2}{|e|B}}
\left[ \frac{2\left(1+u\,e^{-2t}\right)}{2t(1-u)
+u\left(1-e^{-2t}\right)}
-\frac{1+u}{t}
\right].
\label{eq:ut}
\end{equation}
Last, we change to $z=ut \Rightarrow du\, dt = \frac{du\, dz}{u}$ and get
\begin{equation}
\begin{split}
\hat I(L) &= \int_0^{-iq\infty} dz
\int_0^1 \frac{du}{\sqrt{1-u}}\; e^{zq\frac{m^2}{|e|B}}
\left[ \frac{2\left(1+u\,e^{-2z/u}\right)}{2z(1-u)
+u^2\left(1-e^{-2z/u}\right)}
-\frac{1+u}{z} \right]\cr
& \stackrel{q=-1}{=}\int_0^{+i\infty} dz
\int_0^1 \frac{du}{\sqrt{1-u}}\; e^{-z\frac{m^2}{|e|B}}
\left[ \frac{2\left(1+u\,e^{-2z/u}\right)}{2z(1-u)
+u^2\left(1-e^{-2z/u}\right)}
-\frac{1+u}{z} \right].
\end{split}
\label{eq:uz}
\end{equation}

The last operation to perform is a Wick rotation.
$\int_0^{+i\infty} + \int_{1/4\ infinite\ circle} +\int_\infty^0= 2i\pi
\sum residues$. Because of $e^{-z\frac{m^2}{|e|B}}$ the contribution on the
infinite 1/4 circle is vanishing.
That the residue at $z=0$ vanishes is trivial as long as $u$ is not
strictly vanishing. The expansion of the terms between square brackets
in (\ref{eq:uz}) at $z\to 0$ writes indeed
$u-1+ (-\frac53 + \frac{4}{3u} + u) z + \left(-\frac73 - \frac{1}{u^2} +
\frac{7}{3 u} + u\right)z^2+ {\cal O}(z^3)$,
which seemingly displays  poles at $u=0$.
However, without expanding, it also writes, then,
$\frac{2}{2z}-\frac{1}{z}=0$,
which shows that the poles at $u=0$ in the expansion at $z\to 0$
are fake and that the residue at $z=0$ always vanishes.
Other poles (we now consider eq.~(\ref{eq:ut})) can only occur when the
denominator of the first term inside brackets vanishes.
That the corresponding $u_{pole}=\frac{2t}{2t+e^{-2t}-1}$ should
be real constrains them  to occur at $t\to in\pi,n\in{\mathbb N}>0$
and $u\to 1$. In general, they satisfy
$2t(1-u)+u(1-e^{-2t})=0$ which, setting $t=t_1+it_2, t_1,t_2 \in{\mathbb
R}$,
yields the  2 equations $e^{-2t_1}\;\cos 2t_2 = 1+2\eta t_1,\
e^{-2t_1}\;\sin 2t_2 = -2\eta t_2,\ \eta=\frac{1-u}{u}\geq0$.
Since $t_1\to 0$, one may expand the first relation at this limit, which
yields $\cos 2t_2 -1 = 2t_1(\eta+\cos 2t_2)$. As $t_2 \to n\pi$, $\cos 2t_2
>0$ and $\cos 2t_2 -1<0$, which, since $\eta>0$, constrains $t_1$ to stay
negative
\footnote{The 2nd relation then tells us that $\sin 2t_2 <0$, which means
that the poles correspond to $t_2=n\pi-\epsilon,\epsilon >0$.}
. Therefore, the potentially troublesome poles lie in reality
on the left of the imaginary $t$ axis along which the integration is done
and should not be accounted for when doing a Wick rotation. 
After changing  $u$ into $v$ to work from now onwards with the
same notation as in \cite{Jancovici} and ease the comparison, one gets
(\ref{eq:I2}).

\section{Demonstration of (\ref{eq:delm0LLL})}
\label{section:delmLLL}

In (\ref{eq:del0}), we go, like before (see (\ref{eq:chvar1})),
 to the variables $u, y=|e|Bsu$ such that
$du\,ds= \frac{du\,dy}{|e|B u}$ and get
\begin{equation}
\hskip -3cm
\begin{split}
\delta m^0_{LLL} &= \frac{\alpha m}{\pi} \int_0^{\infty} dy \int_0^1
\frac{du}{|e|B\,u\sqrt{1-u}}
\Big[\cos \tan^{-1} \frac{y(1-u)}{u+iy(1-u)}\Big]
\frac{i|e|Bu}{u+iy(1-u)}\;e^{\displaystyle -iyu m^2/|e|B}\;
e^{\displaystyle -i\tan^{-1} \frac{y(1-u)}{u+iy(1-u)}}+c.t._1\cr
&=\frac{\alpha m}{\pi} \int_0^{\infty} dy \int_0^1
\frac{du}{\sqrt{1-u}}
\Big[\cos \tan^{-1} \frac{y(1-u)}{u+iy(1-u)}\Big]
\frac{i}{u+iy(1-u)}\;e^{\displaystyle -iyu m^2/|e|B}\;
e^{\displaystyle -i\tan^{-1} \frac{y(1-u)}{u+iy(1-u)}}+c.t._1
\end{split}
\end{equation}
Next, we go to $t = iy$. This yields
\begin{equation}
\hskip -1cm
\begin{split}
\delta m^0_{LLL} &= \frac{\alpha m}{\pi}\int_0^{+i\infty} dt \int_0^1
\frac{du}{\sqrt{1-u}}
\Big[\cos \tan^{-1} \frac{-it(1-u)}{u+t(1-u)}\Big]
\frac{1}{u+t(1-u)}\;e^{-tum^2/|e|B}\;
e^{\displaystyle -i\tan^{-1} \frac{-it(1-u)}{u+t(1-u)}}+c.t._1\cr
&= \frac{\alpha m}{\pi}\int_0^{+i\infty} dt \int_0^1
\frac{du}{\sqrt{1-u}}
\Big[\cos \tan^{-1} \frac{-it(1-u)}{u+t(1-u)}\Big]
\frac{1}{u+t(1-u)}\;e^{-tum^2/|e|B}\;
e^{\displaystyle -i\tan^{-1} \frac{-it(1-u)}{u+t(1-u)}}+c.t._1
\end{split}
\end{equation}
Last, as before, we go to $z=ut \Rightarrow du\,dt=\frac{du\,dz}{u}$.
\begin{equation}
\hskip -10mm
\delta m^0_{LLL}=
\frac{\alpha m}{\pi}
\int_0^{i\infty} dz \int_0^1 \frac{du}{\sqrt{1-u}}
\Big[\cos \tan^{-1} \frac{-iz(1-u)}{u^2+z(1-u)}\Big]
\frac{1}{u^2+z(1-u)}\;e^{-zm^2/|e|B}\;
e^{\displaystyle -i\tan^{-1} \frac{-iz(1-u)}{u^2+z(1-u)}}+c.t._1
\end{equation}
One has
\begin{equation}
\tan^{-1}(-ix) = (-i)\tanh^{-1}\,x, \quad \cos(-ix) = \cosh x
\end{equation}
therefore
\begin{equation}
\hskip -10mm
\delta m^0_{LLL}=
\frac{\alpha m}{\pi}
\int_0^{i\infty} dz \int_0^1 \frac{du}{\sqrt{1-u}}
\Big[\cosh \tanh^{-1} \frac{z(1-u)}{u^2+z(1-u)}\Big]
\frac{1}{u^2+z(1-u)}\;e^{-zm^2/|e|B}\;
e^{\displaystyle -\tanh^{-1} \frac{z(1-u)}{u^2+z(1-u)}}+c.t._1
\end{equation}
As long as $m\not=0$, the $e^{-zm^2/|e|B}$ and the
$e^{ -\tanh^{-1} \frac{z(1-u)}{u^2+z(1-u)}}$ ensure the convergence on the
infinite 1/4 circle such that, supposing that no pole in the 1/4 quadrant
causes problems, one may do a Wick rotation, which yields
(\ref{eq:delm0LLL}).

\newpage


\end{document}